\documentclass[twocolumn, fleqn]{doc_class}

\usepackage{lineno}

\usepackage[super, comma, sort&compress]{natbib}

\usepackage{multibib}
\newcites{m}{References}
\bibliographystylem{bib_style}
\newcites{met}{References (continued)}
\bibliographystylemet{bib_style}
\newcites{sup}{References (continued)}
\bibliographystylesup{bib_style}

\usepackage{newtxtext, newtxmath}

\usepackage[T1]{fontenc}
\usepackage[utf8]{inputenc}

\usepackage{ae, aecompl}
\usepackage{graphicx}
\usepackage[dvipsnames, hyperref]{xcolor}
\usepackage{xspace}
\usepackage{soul}
\usepackage{booktabs}
\usepackage{adjustbox}
\usepackage{comment}
\usepackage{float}

\usepackage{amsmath}
\usepackage{commath}
\usepackage{siunitx}
\usepackage{nth}

\usepackage[capitalise, noabbrev]{cleveref}
\usepackage{enumitem}
\usepackage{csvsimple}
\usepackage{multirow}
\usepackage{array}
\newcolumntype{P}[1]{>{\centering\arraybackslash}p{#1}}

\crefname{figure}{Fig.}{Figs.}
\crefname{equation}{equation}{equations}

\usepackage[leftmargin=0pt,rightmargin=0pt,skipabove=2pt,skipbelow=2pt,innerleftmargin=5pt,innerrightmargin=5pt,innertopmargin=5pt,innerbottommargin=5pt,linewidth=0pt,innerlinewidth=0pt,middlelinewidth=0pt,outerlinewidth=0pt,roundcorner=0pt]{mdframed}

\newcommand{\orcidsymb}[2]{#1\href{http://orcid.org/#2}{\adjustbox{trim={-.15\width} {0\height} {-.15\width} {0\height},clip}{\includegraphics[height=10pt]{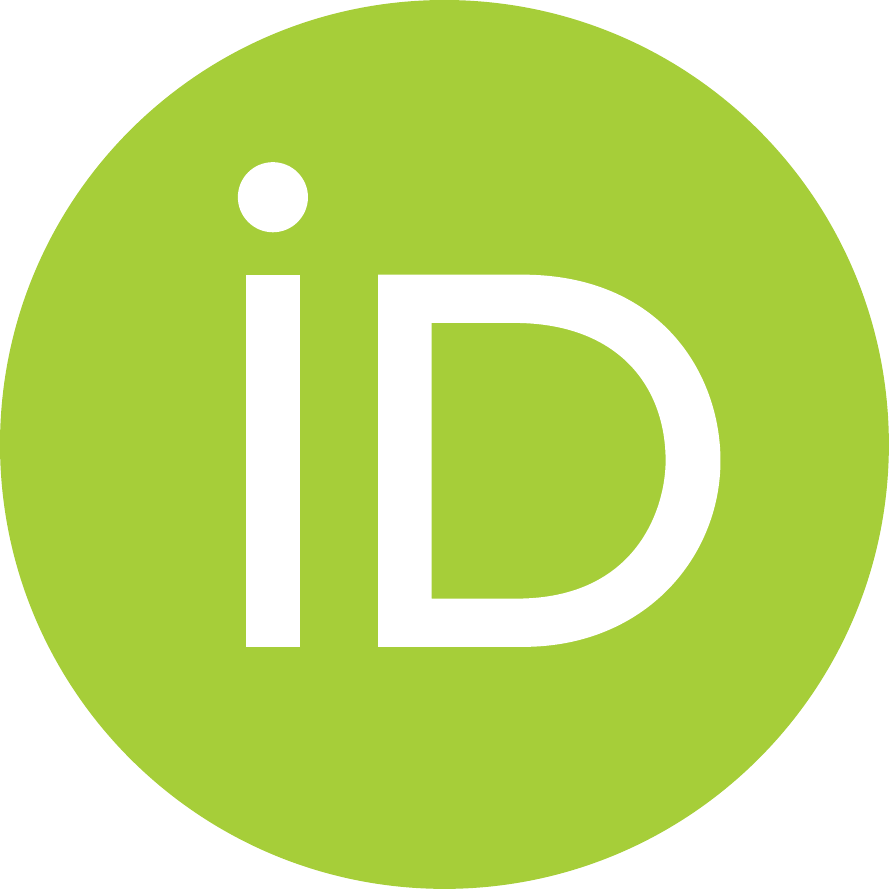}}}}

\newcommand{\Halpha}{\ensuremath{\mathrm{H}\upalpha}\xspace}
\newcommand{\Hbeta}{\ensuremath{\mathrm{H}\upbeta}\xspace}

\hypersetup{
    unicode=true,
    final=true,
    plainpages=false,
    pdfstartview=FitV,
    pdftoolbar=false,
    pdfmenubar=true,
    bookmarksopen=true,
    bookmarksnumbered=true,
    breaklinks=true,
    linktocpage,
    colorlinks=true,
    linkcolor=Blue,
    urlcolor=Blue,
    citecolor=Blue,
    anchorcolor=green
}
\AtBeginDocument{
    \hypersetup{
        pdftitle={A direct black hole mass measurement at the Epoch of Reionization},
        pdfauthor={I. Juodžbalis et al.},
        pdfsubject={},
        pdfkeywords={{galaxies: high-redshift} -- {dark ages, reionization, first stars} -- {methods: observational} -- {techniques: spectroscopic}}
    }
}

\usepackage{paralist}
\makeatletter
    \def\@biblabel#1{\@ifnotempty{#1}{#1}}
    \def\NAT@anchor#1#2{
        \hfilneg\hyper@natanchorstart{#1\@extra@b@citeb}
        #2.
        \hyper@natanchorend
    }
\makeatother

\renewenvironment{thebibliography}[1]{%
    \newpage
    \section*{\normalsize \refname}
    \setlength{\parindent}{0pt}
    \inparaenum
}{\endinparaenum}

\DeclareRobustCommand{\VAN}[3]{#2}
\let\VANthebibliography\thebibliography
\def\thebibliography{\DeclareRobustCommand{\VAN}[3]{##3}\VANthebibliography}

\title[Direct black hole mass measurement]{A direct black hole mass measurement in a Little Red Dot at the Epoch of Reionization}

\author[]{{\orcidsymb{Ignas Juod\v{z}balis}{0009-0003-7423-8660}$^{\hyperlink{inst:Kavli}{1},
\hyperlink{inst:Cav}{2}}$\thanks{E-mail: \href{mailto:ij284@cam.ac.uk}{ij284@cam.ac.uk}},
    \orcidsymb{Cosimo Marconcini}{0000-0002-3194-5416}$^{\hyperlink{inst:UniFlorence}{3}, \hyperlink{inst:Arcetri}{4}}$,
\orcidsymb{Francesco D'Eugenio}{0000-0003-2388-8172}$^{\hyperlink{inst:Kavli}{1}, \hyperlink{inst:Cav}{2}}$,
\orcidsymb{Roberto Maiolino}{0000-0002-4985-3819}$^{\hyperlink{inst:Kavli}{1}, \hyperlink{inst:Cav}{2}, \hyperlink{inst:UCL}{5}}$,}
    \newauthor{
    \orcidsymb{Alessandro Marconi}{0000-0002-9889-4238}$^{\hyperlink{inst:UniFlorence}{3}, \hyperlink{inst:Arcetri}{4}}$,
    \orcidsymb{Hannah \"{U}bler}{0000-0003-4891-0794}$^{\hyperlink{inst:MPE}{6}}$, 
\orcidsymb{Jan Scholtz}{0000-0002-8674-9068}$^{\hyperlink{inst:Kavli}{1}, \hyperlink{inst:Cav}{2}}$,
    \orcidsymb{Xihan Ji}{0000-0002-1660-9502}$^{\hyperlink{inst:Kavli}{1},
\hyperlink{inst:Cav}{2}}$,
\orcidsymb{Santiago Arribas}{0000-0001-7997-1640}$^{\hyperlink{inst:CAB}{7}}$,}
\newauthor{
    \orcidsymb{Jake S.\ Bennett}{0000-0002-8573-2993}$^{\hyperlink{inst:CfA}{8}}$,
    \orcidsymb{Volker Bromm}{0000-0003-0212-2979}$^{\hyperlink{inst:Texas}{9}}$,
   \orcidsymb{Andrew J.\ Bunker}{0000-0002-8651-9879}$^{\hyperlink{inst:Oxford}{10}}$,
   \orcidsymb{Stefano Carniani}{0000-0002-6719-380X}$^{\hyperlink{inst:SNS}{11}}$,
   \orcidsymb{St\'ephane Charlot}{0000-0003-3458-2275}$^{\hyperlink{inst:IAP}{12}}$,}
\newauthor{
    \orcidsymb{Giovanni Cresci}{0000-0002-5281-1417}$^{\hyperlink{inst:Arcetri}{4}}$,
    \orcidsymb{Pratika Dayal}{0000-0001-8460-1564}$^{\hyperlink{inst:Kapteyn}{13, 14}}$,
    \orcidsymb{Eiichi Egami}{0000-0003-1344-9475}$^{\hyperlink{inst:Steward}{15}}$,
    \orcidsymb{Andrew Fabian}{0000-0002-9378-4072}$^{\hyperlink{inst:IoA}{16}}$,
    \orcidsymb{Kohei Inayoshi}{0000-0001-9840-4959}$^{\hyperlink{inst:KIAA}{17}}$,
}
\newauthor{
\orcidsymb{Yuki Isobe}{0000-0001-7730-8634}$^{\hyperlink{inst:Kavli}{1}, \hyperlink{inst:Cav}{2}, \hyperlink{inst:Waseda}{18}}$,
\orcidsymb{Lucy R. Ivey}{0009-0002-5105-1222}$^{\hyperlink{inst:Kavli}{1}, \hyperlink{inst:Cav}{2}}$,
\orcidsymb{Gareth C. Jones}{0000-0002-0267-9024}$^{\hyperlink{inst:Kavli}{1}, \hyperlink{inst:Cav}{2}}$,
\orcidsymb{Sophie Koudmani}{0000-0002-1528-5091}$^{\hyperlink{inst:Herts}{19}, \hyperlink{inst:Catharine}{20}}$,
\orcidsymb{Nicolas Laporte}{0000-0001-7459-6335}$^{\hyperlink{inst:Marseille}{21}}$,
}
\newauthor{
\orcidsymb{Boyuan Liu}{0000-0002-4966-7450}$^{\hyperlink{inst:Heidelberg}{22}}$,
\orcidsymb{Jianwei Lyu}{0000-0002-6221-1829}$^{\hyperlink{inst:Steward}{15}}$,
\orcidsymb{Giovanni Mazzolari}{0009-0005-7383-6655}$^{\hyperlink{inst:MPE}{6}}$,
\orcidsymb{Stephanie Monty}{0000-0002-9225-5822}$^{\hyperlink{inst:IoA}{16}}$,
\orcidsymb{Eleonora Parlanti}{0000-0002-7392-7814}$^{\hyperlink{inst:SNS}{11}}$,}
\newauthor{
\orcidsymb{Pablo G. P\'erez-Gonz\'alez}{0000-0003-4528-5639}$^{\hyperlink{inst:CAB}{7}}$,
\orcidsymb{Michele Perna}{0000-0002-0362-5941}$^{\hyperlink{inst:CAB}{7}}$,
\orcidsymb{Brant Robertson}{0000-0002-4271-0364}$^{\hyperlink{inst:SCruz}{23}}$,
\orcidsymb{Raffaella Schneider}{0000-0001-9317-2888}$^{\hyperlink{inst:Sapienza}{24}}$,}
\newauthor{
\orcidsymb{Debora Sijacki}{0000-0002-3459-0438}$^{\hyperlink{inst:Kavli}{1}, \hyperlink{inst:IoA}{16}}$,
\orcidsymb{Sandro Tacchella}{0000-0002-8224-4505}$^{\hyperlink{inst:Kavli}{1},
\hyperlink{inst:Cav}{2}}$,
\orcidsymb{Alessandro Trinca}{0000-0002-1899-4360}$^{\hyperlink{inst:Como}{25}}$,
\orcidsymb{Rosa Valiante}{0000-0003-3050-1765}$^{\hyperlink{inst:OAR}{26}}$,
\orcidsymb{Marta Volonteri}{0000-0002-3216-1322}$^{\hyperlink{inst:IAP}{12}}$,}
\newauthor{
\orcidsymb{Joris Witstok}{0000-0002-7595-121X}$^{\hyperlink{inst:DAWN}{27},\hyperlink{inst:NBI}{28}}$,
\orcidsymb{Saiyang Zhang}{0000-0003-1541-177X}$^{\hyperlink{inst:Texas2}{29}, \hyperlink{inst:Weinberg}{30}}$
}
}

\pubyear{2025}

\begin{document}
\maketitle

\begin{abstract}
    \begin{mdframed}[backgroundcolor=black!5]
        Recent discoveries of faint active galactic nuclei (AGN) at the redshift frontier have revealed a plethora of broad \Halpha emitters with optically red continua, named Little Red Dots (LRDs) \citem{Matthee2024}, which comprise 15-30\% of the high redshift broad line AGN population \citem{Hainline2025}. Due to their peculiar spectral properties \citem{Ma2025, Ji2025, Juodzbalis2024b} and X-ray weakness \citem{Maiolino_xray_weak}, modeling LRDs with standard AGN templates has proven challenging. In particular, the validity of single-epoch virial mass estimates in determining the black hole (BH) masses of LRDs has been called into question, with some models claiming that masses might be overestimated by up to 
        2 orders of magnitude \citem{Rusakov2025,Naidu2025,Lambrides2024,Lupi2024}, and other models claiming that LRDs may be entirely stellar in nature \citem{Baggen2024}. We report the direct, dynamical BH mass measurement in a strongly lensed LRD at $z = 7.04$. The combination of lensing with deep spectroscopic data reveals a rotation curve that is inconsistent with a nuclear star cluster, yet can be well explained by Keplerian rotation around a point mass of 50 million Solar masses, consistent with virial BH mass estimates from the Balmer lines. The Keplerian rotation leaves little room for any stellar component in a host galaxy, as we conservatively infer $M_{\rm BH}/M_{*}>2$. Such a ``naked'' black hole, together with its near-pristine environment\citem{Maiolino2025}, indicates that this LRD is a massive black hole seed caught in its earliest accretion phase.
    \end{mdframed}
\end{abstract}
\nokeywords




Abell2744-QSO1 (hereafter QSO1) is a strongly lensed, triply imaged system, first discovered in JWST NIRCam imaging by Ref.\citem{Furtak2023}, whose redshift was later confirmed to be $z=7.04$ through NIRSpec prism spectroscopy \citem{Furtak2023_AGN,Greene2024}, which also revealed broad components in \Halpha and \Hbeta lines. The compactness, red optical (rest-frame) slope together with blue UV (rest-frame) slope, classify it as a typical ``Little Red Dot'' (LRD) \citem{Kocevski2024,Hainline2025}. Further observations of QSO1 using the high resolution R2700 NIRSpec Integral Field Spectroscopy (IFS) mode \citem{Ji2025, DEugenio2025} clearly spectrally resolved the broad- and narrow-line emission, as well as detected line variability \citem{Ji2025, Furtak2025variab}, thereby robustly identifying QSO1 as hosting an accreting black hole (BH).
Based on virial relations using broad line widths and luminosities, a black hole mass of about $4\times 10^7~M_\odot$ was estimated\citem{Furtak2023_AGN}.
In addition, recent work \citem{Ji2025, DEugenio2025} has established a high BH to dynamical mass ratio for QSO1 ($M_{\rm BH}/M_{\rm dyn} > 0.2$). However, these results rest on the assumption that ``virial relations'' \citem{VolonteriBHmass} that are calibrated locally, are still applicable at z=7. While local virial relations have been tested out to $z\sim2.3$ \citem{Abuter2024} for luminous quasars (and found adequate for H$\alpha$ within a factor of 2.5), there are currently no validating measurements beyond z>2, and for the LRD class of objects. In this work, we provide the first direct BH mass measurement in the high redshift Universe, indeed illustrating that virial BH mass calibrations apply to this prototypical LRD.

It is first interesting to note that, given the low narrow line velocity dispersion in QSO1\citem{DEugenio2025} ($\sigma _N=22$~km~s$^{-1}$, see also Methods), the black hole's sphere of influence, assuming the mass estimated from the virial relations, has a radius of $\sim 270$~pc, which should be resolvable with JWST, also thanks to the gravitational lens shearing (a factor of $\sim$3.5\citem{Furtak2023_AGN}). However, we will not make any a priori assumptions about the black hole mass in the following analysis.

\begin{figure*}
    \centering
    \includegraphics[width=\linewidth]{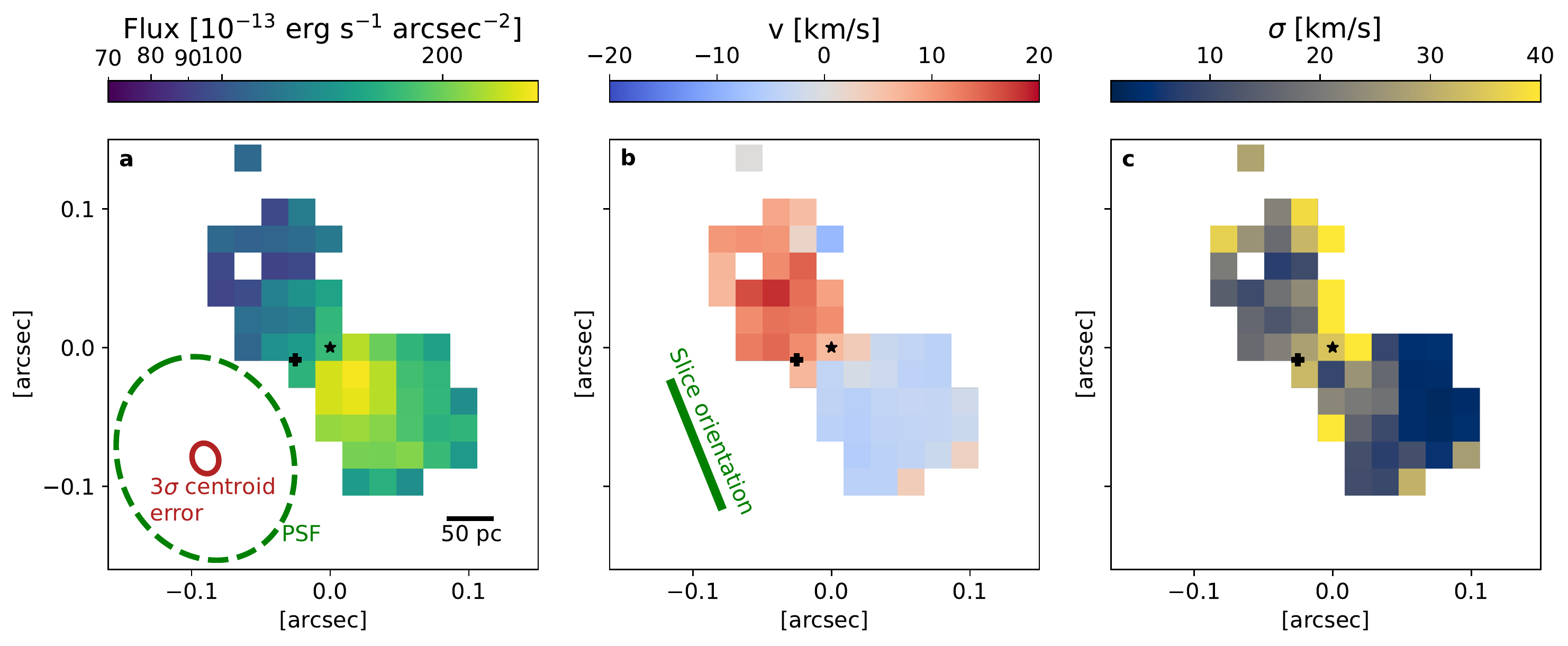}
    \caption{\textbf{H$\alpha$ narrow line emission and kinematics maps of QSO1.} Panel \textbf{a} shows the narrow line surface flux, where only spaxels with narrow H$\alpha$ S/N $>5\sigma$ included in the map. The dashed green ellipse shows the dimensions of the PSF at the wavelength of H$\alpha$. The smaller red ellipse shows the 3$\sigma$ position accuracy at our S/N and pixel scale. The black segment shows the image scale in parsecs. Panel \textbf{b} shows the line velocity map, while panel \textbf{c} shows the line velocity dispersion $\sigma_v$. The black star indicates the position of the dynamical center from \texttt{MOKA3D} fitting. The position angle (PA) of the IFU slicer is shown by the green line, indicating that the observed velocity gradient is not due to calibration artefacts. The centroid of the broad line emission is indicated with a black cross.}
    \label{fig:rot_curve}
\end{figure*}

We perform a detailed analysis of the spatial and kinematic information present in the current deep, high resolution NIRSpec IFS data, resampled to a 0.02'' pixel scale, by focusing on kinematical maps of the narrow \Halpha emission. These maps were obtained by subtracting the broad line and continuum emission from the data cube and fitting the resulting cube containing only the narrow \Halpha emission. The details of this procedure are given in the Methods. The line flux, velocity, and dispersion maps are shown in Fig \ref{fig:rot_curve}. As can be seen in Fig. \ref{fig:rot_curve}b, the narrow H$\alpha$ velocity field reveals an apparent velocity gradient  with an amplitude of $\sim  10~km~s^{-1}$.

\begin{figure*}
    \centering
\includegraphics[width=0.55\linewidth]{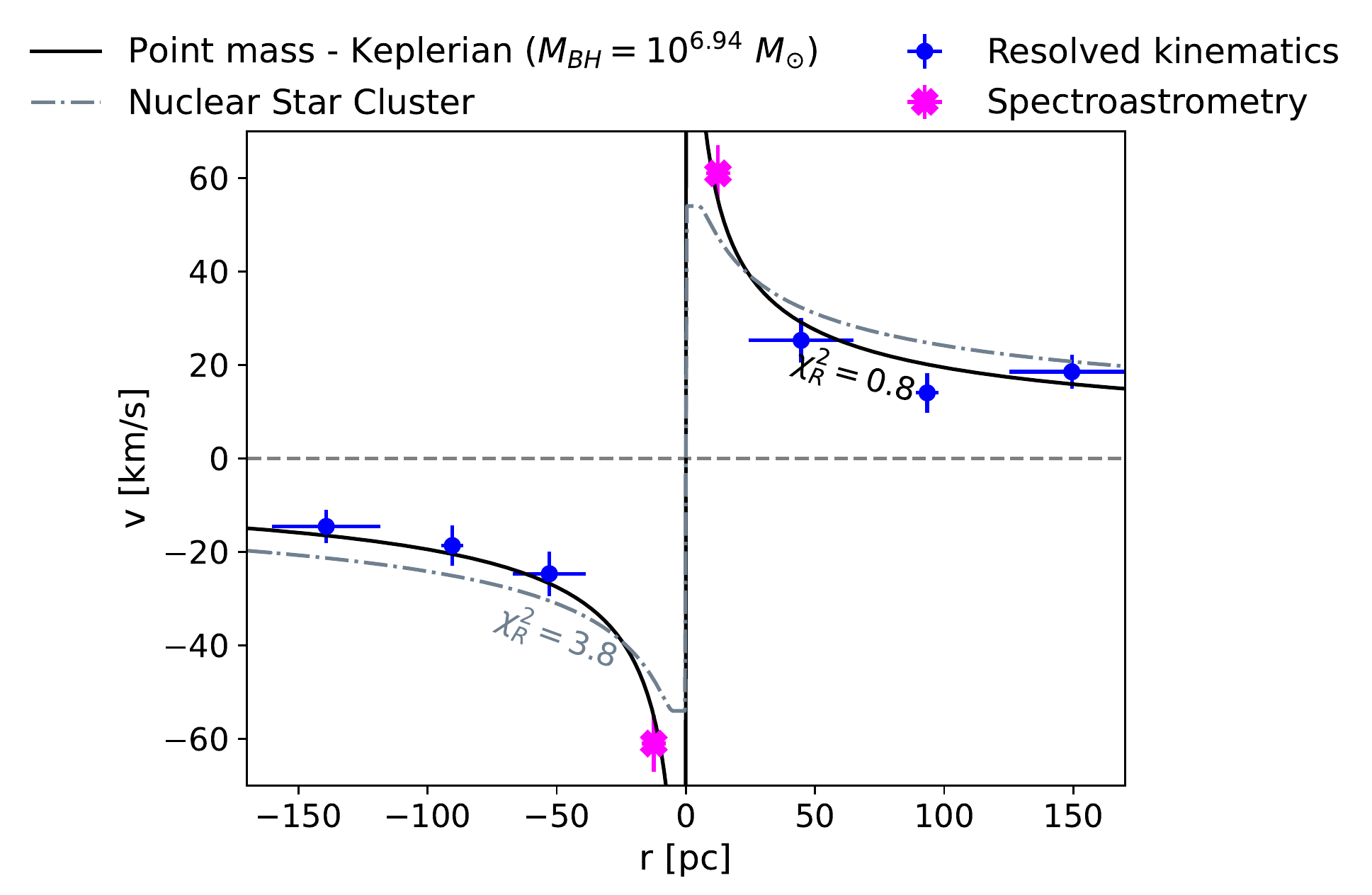}
\includegraphics[width=0.9\linewidth]{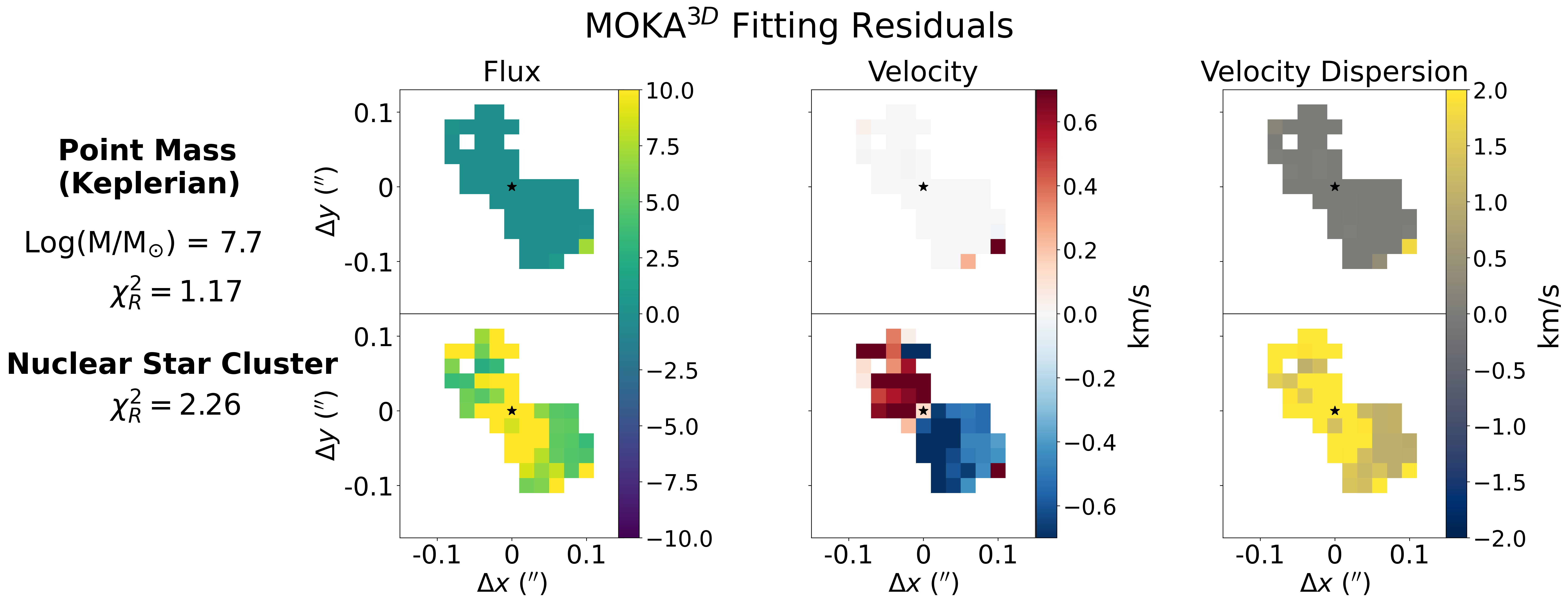}
    \caption{\textbf{Direct dynamical measurements of the black hole mass.} \textbf{Top Panel: } 1D rotation curve along the lines of nodes. Blue points are from the binned 2D velocity field, while the magenta crosses are the spectro-astrometry measurements. The solid black line indicates the Keplerian best fit with a point mass, giving a BH mass of $8.3\times 10^6~M_\odot$ (which is a lower limit given that with this method the inclination is not constrained). The dot-dashed line is for a Nuclear Star Cluster (see text), which results in a worse fit. \textbf{Bottom Panel: }Comparisons of the residuals of the full MOKA3D fits. The point mass Keplerian rotation gives residuals far smaller than the Nuclear Star Cluster model.}
    \label{fig:BHmeas}
\end{figure*}

The narrow line emission is spatially extended on scales of up to $\sim$200~pc, as found by Ref.\citem{Maiolino2025} (note that the continuum is instead unresolved, as discussed in Ref.\citem{Furtak2023_AGN}). We sample the rotation curve by averaging the positions of positive and negative velocity spaxels across three spatial bins corresponding to 50, 100 and 150~pc distances from the rotation centre (see Methods). The node velocities are obtained through an inverse variance weighted mean of the contributing spaxels, corrected for velocity dispersion support by adding $\sigma _v$ in quadrature. The binned velocities are shown as blue circles in Fig. \ref{fig:BHmeas}-top.

While the narrow line kinematics provide useful constraints on rotation on $\sim 100$~pc scales, the small size of the object, coupled with the comparatively low spatial resolution,
makes it difficult to trace the inner parts of the resolved rotation curve. However, given the high signal-to-noise on H$\alpha$, it is possible to recover kinematic information below the spatial resolution beam. Specifically, we employ spectroastrometric techniques, to search for rotation in the inner part of the narrow line region (note that the same analysis cannot be performed using H$\beta$ narrow or [OIII], as these lines are too weak). Briefly, spectroastrometry consists in identifying the centroid positions when scanning spectroscopic channels of a line \citem{Gnerucci2011}, producing a map of average gas positions across a velocity range, a technique previously used by GRAVITY+ to measure BLR sizes and BH masses in QSOs\citem{Abuter2024}. The full description of the spectroastrometric analysis is given in the Methods. We find that the centroids of the \Halpha images in +50~km~s$^{-1}$ and -50~km~s$^{-1}$ velocity channels are separated by $24.9\pm 9.4$~pc in the source plane (Extended Data Fig. \ref{fig:spectro_centroids}). Taking half of this separation gives a radius scale $r_{\rm spec} = 12.5_{-4.7}^{+4.7}$~pc, which, coupled with $\mathrm{FWHM}=52_{-14}^{+14}$~km~s$^{-1}$ \citem{DEugenio2025},  yields an enclosed mass of $\log{(M_{\rm spec}/\mathrm{M}_{\odot})} = 6.90_{-0.23}^{+0.23}$. It should be noted that this estimate is a lower limit, as the inclination is unconstrained with this method.

The spectroastrometric measurements above indicate a compact and dense system, however, they alone cannot exclude a significant contribution of stars, gas or dark matter to the mass budget. To get a more complete picture of the mass distribution in our object, we combine the spectroastrometric measurements with the large scale rotation in order to construct a rotation curve for QSO1 (Fig. \ref{fig:BHmeas}-top). In order to test if this rotation has significant contribution from an extended stellar component, we fit the data in Fig. \ref{fig:BHmeas}-top with two models - (i) a point mass and (ii) a compact, yet extended, mass distribution mimicking the (thoroughly studied) nuclear star cluster (NSC) of the Milky Way, with a density profile of a broken power law where the transition between $\rho \propto r^{-2}$ and $\rho \propto r^{-3}$ occurs at $R_c = 5$~pc \citem{Schodel2014}. The details of the fitting and the exact functional form are given in the Methods, however Fig. \ref{fig:BHmeas}-top shows that an extended MW-like NSC mass distribution is disfavored ($\chi^2_R=3.8$) when compared to a pure point mass (Keplerian) model ($\chi^2_R=0.8$), with $\log{M_{BH}/M_{\odot}} = 6.94 \pm 0.15$. The evidence from kinematics alone corresponds to $>5\sigma$ preference for a point mass. 
Additional kinematic evidence ruling out the NSC (as well as more extensive NSC cases) will be presented further below and in the Methods, also through the full 2D kinematic fitting.
Here we note that the implied NSC masses, $\log{M_*(<R_c)} = 6.52$ and $\log{M_*(<100 \, {\rm pc})} = 7.1$, are considerably above the stellar mass limits derived from the mass-to-light ratio of the object (see Methods).

We also note that in these tests the NSC effective radius $R_c$ was fixed to $5$~pc, as in the MW's NSC -- if left free, then the $R_c$ collapses to $10^{-4}$~pc, hence effectively mimicking a point-mass. In the Methods we estimate an upper limit on $R_c$ of 0.2~pc, which would make the putative NSC in QSO1 more than 1~dex more concentrated than the densest nuclear clusters found in the local Universe, and also more concentrated than the densest star clusters found in the early universe (Extended Data Fig. \ref{fig:nsc_max_extent}).

In addition to the NSC, we also test the Plummer sphere model \citem{Plummer1911}, which typically describes Globular Clusters, and which posits a near constant density out to a certain scale radius $R_p$ and an $\rho \propto r^{-5}$ drop off once the scale radius is exceeded. This model reproduces the observed 1-dimensional rotation curve well, however, the best-fit $R_p \sim 10^{-4}$~pc and mass $\log (M_0/M_\odot) \sim 6.9$ effectively reduce this model to a point mass (see Methods). In the Methods we also test the scenario of a nuclear Dark Matter cusp, which runs into the same problems as the NSC and Plummer sphere.

Therefore, the simplest and most direct interpretation of the rotation curve is a central point mass of $\log{M_{BH}/M_{\odot}} = 6.94 \pm 0.15$, consistent with the bare spectroastrometric estimate, and corresponding to a supermassive black hole. Once again, as the inclination of the rotation is unconstrained by this method, this mass estimate is a lower limit.

In order to examine the robustness of our conclusions against sources of systematic uncertainty, and without having to rely on spectroastrometry, we re-analyze our data by constructing self-consistent kinematics models by using the \texttt{MOKA3D} framework \citem{Marconcini2023, Marconcini2025}, which takes into account the detailed flux distribution of the kinematic tracers, inclination effects, and the point spread function. The mass distribution models considered are the same as in direct fitting above - the point mass, NSC and Plummer sphere. The residuals of the fitting models for the point mass and NSC are shown in the bottom panel of Fig.\ref{fig:BHmeas} (while the Plummer sphere is shown in the Methods). Through this independent analysis, we find that the best fitting model ($\chi^2_R=1.17$) is Keplerian rotation around a point mass of $\log{M} = 7.7 \pm 0.3$, when corrected for $52 \pm 2^{\circ}$ inclination estimated via the same method (see Methods), which is consistent with the lower limit obtained above from the direct fits to the rotation curve. The Plummer sphere model can also provide a good fit ($\chi^2_R = 1.60$), however, it does so by collapsing the sphere to a point mass with a best-fit $R_p = 0 ^{+3}_{-0}$~pc (see Methods), just as for the direct 1D rotation fitting discussed above. The NSC model does remain extended ($R_c = 4 \pm 2$~pc) when fitted to the full 2D kinematics, however, the considerable systematic residuals, as illustrated in Fig. \ref{fig:BHmeas}-bottom, and the much higher $\chi^2_R=2.26$, indicate that this model is an inadequate fit to the data. Finally, we also robustly exclude kinematic contamination from outflows utilizing a combination of spectroastrometry and \texttt{MOKA3D} modeling (see Methods for details).

\begin{figure*}
    \centering
    \includegraphics[width=0.9\textwidth]{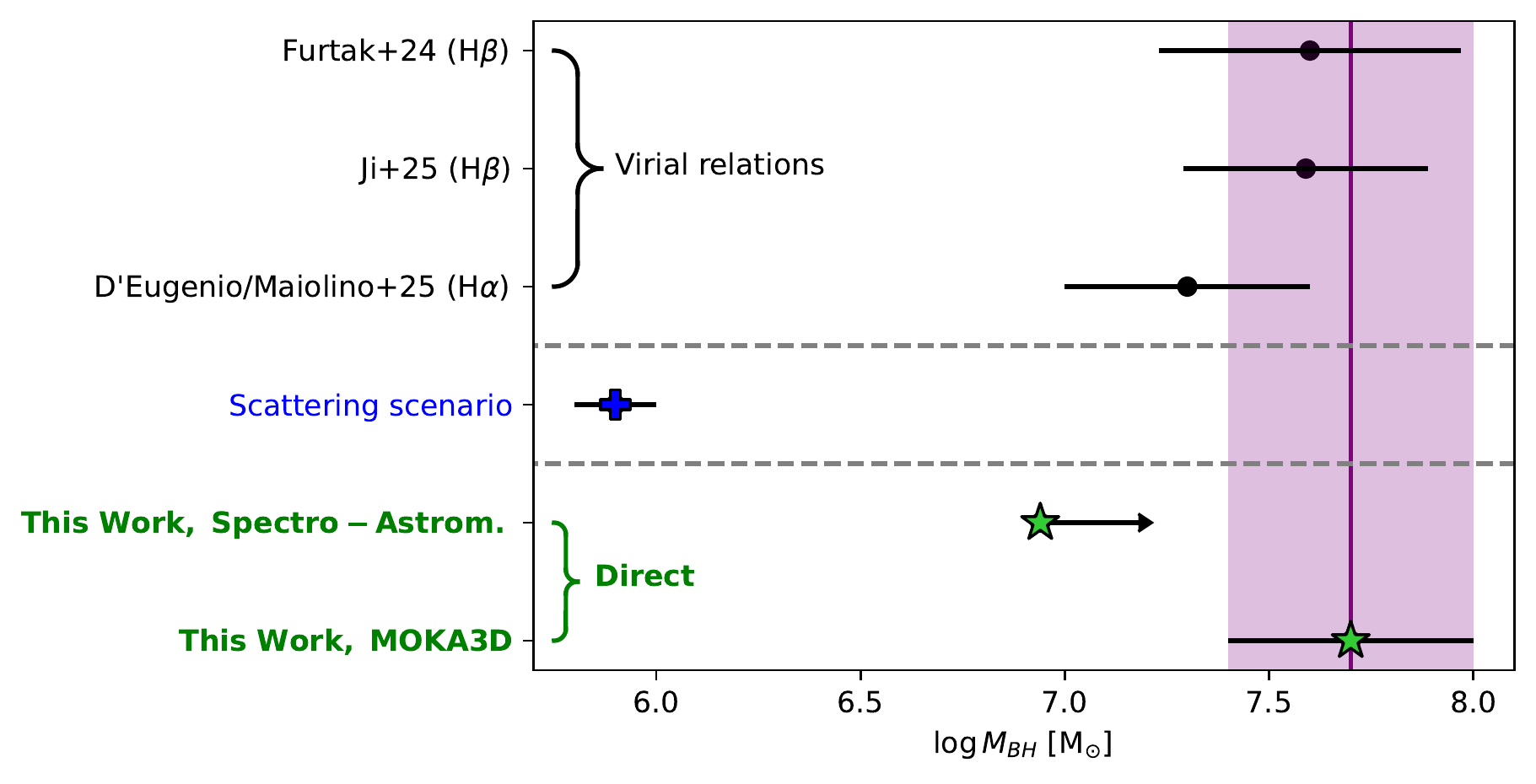}
    \caption{\textbf{Summary of BH mass estimates for QSO1. }Comparison between our lower limit and \texttt{MOKA3D} direct measurements (green stars) with previous virial estimates (black circles) \protect\citem{Furtak2023_AGN,Ji2025,DEugenio2025,Maiolino2025} and the scenarios assuming that the dominant broadening mechanism of the broad lines is due to electron scattering \protect\citem{Rusakov2025}. The purple shaded region shows the 1$\sigma$ uncertainty on the MOKA3D estimate.}
    \label{fig:mbh_comp}
\end{figure*}

These results represent the first direct, dynamical measurement of a BH mass at the Epoch of Reionization and in an LRD. An immediate implication is that alternative scenarios explaining this LRD without accreting black holes are essentially ruled out. Additionally,
we can investigate the reliability of single-epoch BH mass virial estimates out to the Epoch of Reionization, and specifically for LRDs. Fig. \ref{fig:mbh_comp} shows a comparison between our BH mass direct measurement and other literature estimates obtained using the virial scaling relations. As can be seen, our spectroastrometry lower limit and the MOKA3D measurement are  entirely consistent with these estimates. On the other hand, an electron scattering scenario \citem{Rusakov2025}, while providing a good spectral fit to the broad lines (see Extended Data Fig. \ref{fig:scattering}), underestimates the BH mass by nearly 2~dex (Fig.\ref{fig:mbh_comp}). Other scenarios, such as Balmer scattering, which have been claimed to also give black hole masses two orders of magnitude lower\citem{Naidu2025} relative to the virial relations, are also ruled out.

The resulting Eddington luminosity of the black hole is $7.6\times 10^{45}~$erg~s$^{-1}$. By using standard scaling relations between broad H$\alpha$ and bolometric luminosity\citem{SternLbol} (and using the extinction $A_V=0.66\pm0.40$ derived by Ref.\citem{Maiolino2025}), we infer that the black hole is accreting well below its Eddington limit, with $L/L_\mathrm{Edd}\approx 0.03$. If the broad H$\alpha$ to bolometric luminosity relation is higher than estimated locally, as suggested by Ref.\citem{Maiolino_xray_weak}, then this would imply $L/L_\mathrm{Edd}\sim 0.01-0.02$, indicating that
the black hole might be in a near-dormant state. Yet, the black hole may still have experienced previous super-Eddington bursts, as inferred by Ref.\citem{Juodzbalis2024} for another over-massive, dormant black hole at a similar redshift.

\begin{figure*}
    \centering
    \includegraphics[width=\textwidth]{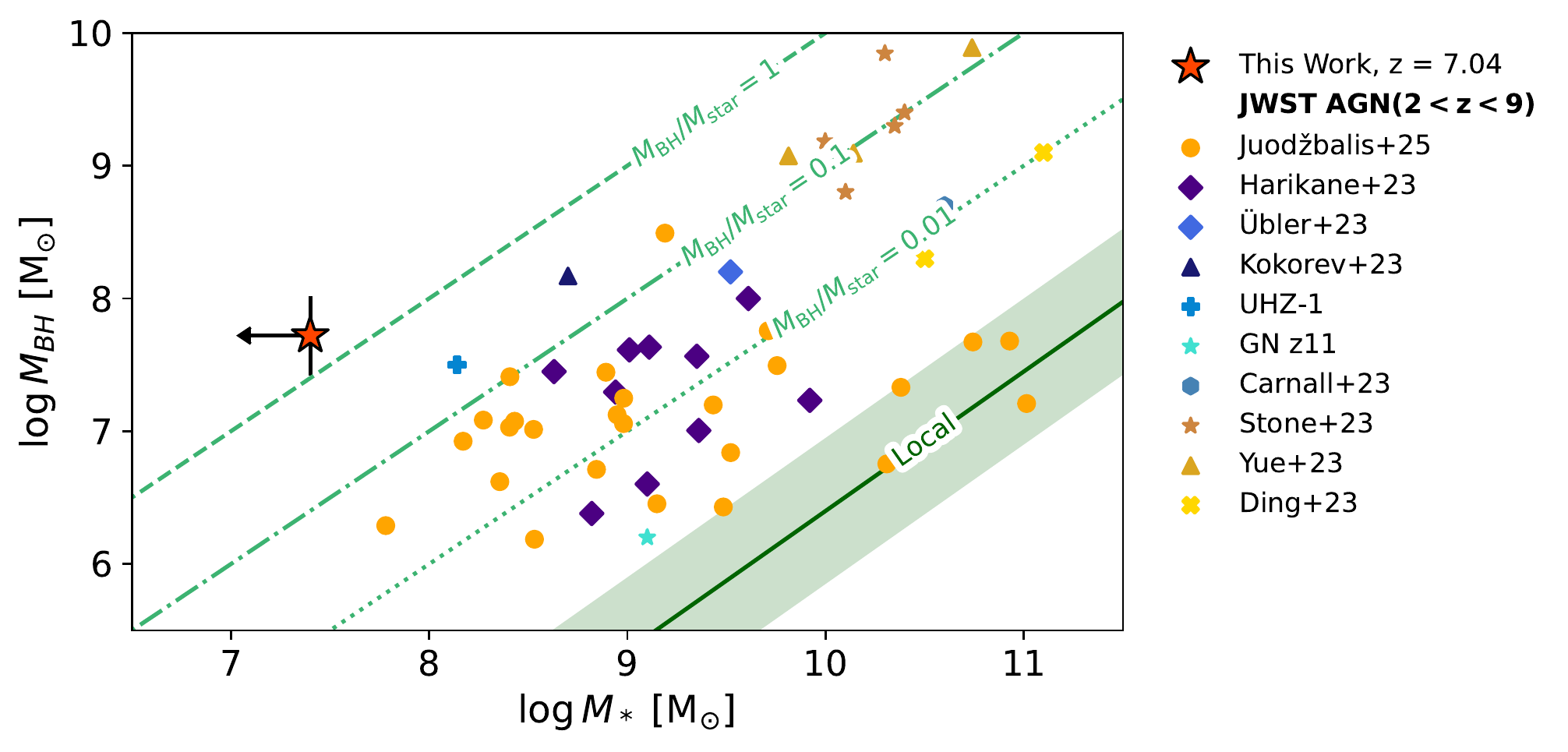}
    \caption{\textbf{Location of QSO1 on the $M_{BH} - M_{*}$ plane.} QSO1 is indicated by the red star. The remaining points represent measurements from other JWST observations of low mass AGN \protect\citem{Juodzbalis2025, Harikane_AGN, Ubler2023, Kokorev2023_AGN, Goulding2023_AGN, Maiolino24_GN-z11, Carnall2023} and quasars \protect\citem{Yue2024_QSO, Stone2023_QSO, Ding2023_QSO}. The solid green line shows the local scaling relation from Ref. \protect\citem{VolonteriBHmass}, with the scatter indicated by shading. The other green lines indicate constant $M_{BH}/M_{*}$ ratios. With $M_{BH}/M_{*} > 2$, QSO1 lies orders of magnitude above the local scaling relations and  is $\sim1$~dex more overmassive than even the most extreme sources found by JWST so far.}
    \label{fig:mbh_mstar}
\end{figure*}

Finally, we note that the Keplerian rotation curve leaves little room for any stellar component. Specifically, in the Methods we derive a dynamical upper limit  on the stellar mass in the host galaxy of $M_*<2\times 10^7~M_\odot$. This is very conservative as it is assuming an exponential disc distribution, and it includes any distributed mass components such as dark matter and gas (see Methods). As discussed in the Methods, a powerlaw profile with $\rho\propto r^{-3}$ or a NSC-like mass distribution gives a similar upper limit. This upper limit makes QSO1 the most ``naked'' massive black hole ever found, with $M_{\rm BH}/M_*>2$, and in line with the previous finding that this black hole is hosted in an environment that is chemically nearly pristine \citem{Maiolino2025}.
This demonstrates the possibility of black-hole primacy, i.e. black holes forming and growing earlier and/or much faster than their host galaxy.
The lower limit on the $M_{\rm BH}/M_*$ ratio is three orders of magnitude higher than observed locally. Fig.\ref{fig:mbh_mstar} shows how extreme  QSO1 is on the $M_{\rm BH}$ versus $M_*$ diagram relative the local relation - located at the extreme tail of the overmassive black holes identified by JWST in previous surveys\citem{Juodzbalis2025}.

The only scenarios that can account for such a system are those invoking ``heavy seeds'', such as Direct Collapse Black Holes (DCBHs, resulting from the direct collapse of massive pristine clouds), or Primordial Black Holes (PBHs, formed in the first second after the Big Bang)\citem{BL2003_DCBH,Natarajan2024,Zhang2025_PBH}.

Yet, most DCBH scenarios would require a strong source of UV radiation in the vicinity, which is not seen (not even a post-starburst galaxy that might have produced UV radiation in the past), although some scenarios may expect direct collapse in different environments\citem{Wise2019}. However, DCBH models suggest that their early growth is limited by the baryon fraction in an atomically-cooling halo\citem{BrommYoshida2011_Review,Pacucci2015_DCBH}, placing an upper limit to the $M_{\rm BH}/M_{\rm dyn}$ ratio of $\sim 0.1$, i.e. more than 1~dex lower than our inferred lower limit. 

On the other hand, some independent evidence for the PBH scenario comes from the very low metallicity of this system, as discussed in Ref\citem{Maiolino2025}. However, the observed mass of $5\times 10^7~M_\odot$ is significantly higher than the preferred $10^6~M_\odot$ PBH  mass scale given by the electron-positron annihilation epoch in the ultra-early universe\citem{Carr2020}; therefore, the observed mass would require either significant accretion or rapid merging of PBHs, which may be linked to their highly clustered nature\citem{Zhang2024_BiasedPBH}.
The intermediate scenario of ``Not-Quite-Primordial Black Holes'' (NQPBHs)\citem{Qin2025_NQPH}, whereby heavy seeds formed at $1+z>200$, when the Cosmic Microwave Backgroud was energetic enough to photo-disocciate molecular hydrogen, hence suppressing cooling, could also be a viable, alternative possibility. 

Regardless of the specific model, the high mass in such a remote cosmic epoch, the extremely high  $M_{\rm BH}/M_*$, together with the near-pristine environment\citem{Maiolino2025}, indicate that QSO1 is a massive black hole seed caught in the earliest phases of accretion\citem{Inayoshi2025_firstBHs}.

\clearpage
\bibliographym{AGN}

\clearpage
\makeatletter
\renewcommand{\fnum@figure}{Extended Data Fig. \thefigure}
\renewcommand{\fnum@table}{Extended Data Table \thetable}
\makeatother
\setcounter{figure}{0}
\setcounter{table}{0}
\setcounter{footnote}{0}


\newpage

\section*{Methods}
\subsection*{Conventions}
Throughout this work we assume a flat $\Lambda$CDM cosmology with $\Omega_m = 0.315$, $H_0 = 67.4$~km~s$^{-1}$~Mpc$^{-1}$ \citemet{Planck2020}. All reported magnitudes are in the AB system. Following the lensing model of \citem{Furtak2023_AGN}, we adopt a flux magnification factor $\mu = 6.2$ and a shear factor of 3.52 for our source (image A of QSO1). Hence, 1 arcsec in the image plane corresponds to 1.52 physical kiloparsecs (pkpc). For robustness tests we use the Bayesian Information Criterion (BIC), defined as $BIC \equiv \chi^2 + k\ln{n}$, where $k$ is the total number of model parameters and $n$ - the number of points fitted, a decrease in BIC, $\Delta \mathrm{BIC} \geq 5$ between two models was required for robust preference of one over the other, although our main conclusions remain unchanged even if a stricter $\Delta \mathrm{BIC} \geq 10$ threshold is adopted.

\subsection*{Data reduction}
We use data from the BlackTHUNDER NIRSpec-IFU survey, focusing on the 7.3-hour exposures with the G395H grating, giving a nominal spectral resolution $R=3,700$ at the wavelength of H$\alpha$ \citem{DEugenio2025}.
The NIRSpec IFU was centered on image A of QSO1 ($RA=00:14:19.161$; $DEC=-30:24:05.664$)\citem{Furtak2023}.
A detailed description of the reduction procedures is available in \citem{Ji2025, DEugenio2025}, however, a summary will be provided here for context.

The spectra were extracted following the procedures of \citemet{Perna+2023}, but using version 1.17.1 of the JWST pipeline. At $z = 7.04$, the H$\alpha$ line falls just outside the nominal wavelength coverage, however, the F290LP filter does not cut off longer wavelengths and the detector efficiency allows to recover H$\alpha$ emission. We perform this recovery by extrapolating the wavelength solution, flat-field curves and the grating equation derived LSF out of the nominal range and towards the detector sensitivity limit of $\lambda = 5.34$~$\mu$m. The peak of the H$\alpha$ line of Abell2700-QSO1 falls on $\lambda = 5.278$~$\mu$m, hence our modification readily recovers the entirety of H$\alpha$ emission. While flux calibrations beyond the nominal range may suffer inaccuracies, the primary interest of this work is a kinematics study, hence our key kinematics results are insensitive to flux calibrations. The BH mass measurements are more affected, however, the square root dependence of BH mass on luminosity means that flux calibrations have to be wrong by an order of magnitude to significantly impact the measurements.

The nominal spaxel scale of the processed data was 0.05", however, utilizing the large number of dithers we are able to oversample the cube to a scale of 0.02" per spaxel without incurring significant sampling artefacts. We choose the 0.02" cube for the main kinematic and spectroastrometric analysis, with the 0.05" cube used to perform consistency checks, ensuring that our results are not pixel sampling artefacts.

\subsection*{Narrow line kinematics}
In order to constrain the narrow line kinematics, we perform spaxel-by-spaxel fitting of the original cube to subtract the broad H$\alpha$ and the continuum components. The broad H$\alpha$ emission was modeled with the same shape as derived from combined spectrum in \citem{DEugenio2025}, consisting of two Gaussians with FWHM of 450 and 1800~km~s$^{-1}$ and containing 41\% and 59\% of the total broad line flux respectively. This composite profile was then scaled in the fitting procedure to reproduce the data in each spaxel. In addition to the broad line emission, the H$\alpha$ profile of Abell2744-QSO1 also includes significant absorption. We account for this in the fitting by including an absorption profile of the form:

\begin{equation}
\label{eq:abs}
    \frac{F_{\lambda}}{F_{\lambda;0}} = 1 - C_f + C_fe^{-\tau_\lambda},
\end{equation}
where $\frac{F_{\lambda}}{F_{\lambda;0}}$ is the ratio of transmitted to emitted fluxes,  $C_f$ is the covering factor and $\tau_{\lambda}$ - a Gaussian optical depth distribution characterized by the optical depth at the core of the line ($\tau_0$), Doppler broadening ($b$) and velocity offset from the narrow emission ($dv$). Following the results of \citem{DEugenio2025}, we adopt $C_f = 0.55$, $\tau_0 = 1.9$, $b = 100$~km~s$^{-1}$ and $dv = -36$~km~s$^{-1}$ as the parameters to \autoref{eq:abs}. The narrow H$\alpha$ line was parametrized as a Gaussian, with $\sigma$ bounded between 10 and 50~km~s$^{-1}$ and peak location allowed to shift by 100~km~s$^{-1}$ relative to the redshift obtained from an aperture spectrum.

After subtracting the broad line and continuum emission, we perform Bayesian fitting of the narrow H$\alpha$ emission. We choose a flat prior on the FWHM of the narrow H$\alpha$ line, bounded between 10 and 200~km~s$^{-1}$, while the redshift prior was a Gaussian centered on $z = 7.0367$ \citem{DEugenio2025, Ji2025} and had a FWHM of 100~km~s$^{-1}$ in velocity space. The posteriors were estimated using a Monte Carlo integrator \texttt{emcee} \citemet{emcee} with only spaxels for which the line was detected at $>5\sigma$ significance considered for further analysis. The H$\alpha$ emission and kinematics maps obtained as the median values of the posteriors are shown in Fig.\ref{fig:rot_curve}. The velocities were calculated from the difference between the \Halpha redshift in each spaxel and the median value.

As can be seen in the figure, the narrow H$\alpha$ kinematics reveal a strong velocity gradient of an amplitude of 10~km~s$^{-1}$. This value is smaller than the spectral pixel scale of 20-25~km~s$^{-1}$, but it is well known that with high signal-to-noise it is possible to measure offsets below the spectral and spatial resolutions \citemet{Ivison2007}, with an accuracy that is given by $\Delta v\propto FWHM_{LSF}\times (S/N)^{-1}$. However, wavelength calibration issues may potentially introduce spurious effects. In particular, due to data-processing issues, point sources may display spurious rotation of up to $\lesssim 1$ spectral pixel (Beck T., in~prep.\footnote{Available on \href{https://jwst-docs.stsci.edu/known-issues-with-jwst-data/nirspec-known-issues/nirspec-ifu-known-issues\#gsc.tab=0}{jdocs.}}). In case of false rotation, the inferred velocity gradient is always across the IFU slicers, because spaxels along the same slice tend to have a constant velocity offset with respect to the centre of the source.
However, in our case, the observed velocity gradient is aligned to the slicers, i.e., orthogonal to the direction expected for potential, spurious rotation. We can therefore rule out the observed rotation being due to data-reduction artifacts.

While the narrow line emission is spatially extended up to 200~pc scales as found by \citem{Maiolino2025}, this extent is not much larger than the IFU PSF at H$\alpha$ wavelengths \citem{DEugenio2025}. As shown in Fig. \ref{fig:rot_curve}, QSO1 is resolved into only two beams. We thus do not attempt to construct a rotation curve through a full line of nodes. Instead, we bin the positions of positive and negative velocity spaxels to map out the red and blue wings of the rotation curve. The bins were chosen to include approximately equal numbers of spaxels at 50, 100 and 150~pc distances from the center of rotation (black star in Fig.\ref{fig:rot_curve} - this was found iteratively as part of the fit, and then confirmed to be consistent with the best fit obtained by MOKA3D). The binned velocities are obtained through an rms weighted mean of the contributing spaxels in each bin. We then take into account pressure support by adding in quadrature the velocity dispersion, $\sigma$, to the observed velocity in each spaxel \citemet{Dalcanton2010} with the final error on spaxel rotational velocity being a combination of the fitting errors on $\sigma$ and $v$ values. When conducting the averaging, we conservatively assume the largest spaxel velocity error to apply to the entire bin. Hence, our velocity errors are $\sim 6$~km~s$^{-1}$, similar to what is reasonably achievable at our resolution and signal-to-noise \citem{DEugenio2025}. It should be noted that using the median redshift as the velocity zero point can skew the observed rotation as the narrow \Halpha flux distribution is skewed towards the blue side (Fig. \ref{fig:rot_curve}). This causes more of the pixels on the blue side to be above the S/N threshold for inclusion into the extended kinematics map, skewing the median to one side. In order to account for this bias, we force the mean velocity across both wings to be zero. Note that in the MOKA3D modelling, discussed below, the determination of the zero velocity is a free parameter of the global fit; the two rest frame central velocities, independently determined with the two methods, are consistent with each other.

\subsection*{Spectroastrometric measurements}
The spatial resolution (0.1") of the NIRSpec IFU data is larger than the 30~pc scales, below which the influence on kinematics from a point mass with a few tens million masses would be most apparent. However, both spatial and spectral information present in the data can be combined to measure sizes significantly below the nominal resolution, if the signal-to-noise is high enough. The method through which this is accomplished, spectroastrometry, was first conceived to measure separations between close binary stars \citemet{Beckers1982, Aime1988} and has been applied to search for BHs in the centers of galaxies by \citem{Gnerucci2011} and, more recently, to measure BLR sizes and BH masses in bright QSOs \citem{Abuter2024}.

The spectroastrometric measurements, when carried out on an IFU cube containing an emission line involve slicing the line into several velocity channels an constructing an image of the source in each. Afterwards, centroids in each velocity channel are measured and distances to the reference zero velocity channel are computed. With sufficient signal to noise, this method can recover a rotation curve at scales smaller than the PSF by leveraging the fact that centroid accuracy for a given S/N and spatial resolution is given by $\sigma_C = 0.6\theta(S/N)^{-1}$ \citemet{Ivison2007}, where $\theta$ is the size of the PSF. Given the peak S/N = 12 of the \Halpha line, our nominal centroid accuracy is thus $\sigma_C \sim 5$~mas, corresponding to 2~pc on the source plane. In practice, the measurement error is larger than this value as it is also influenced by flux errors on individual pixels making up the image.

\begin{figure}
    \centering
    \includegraphics[width=\linewidth]{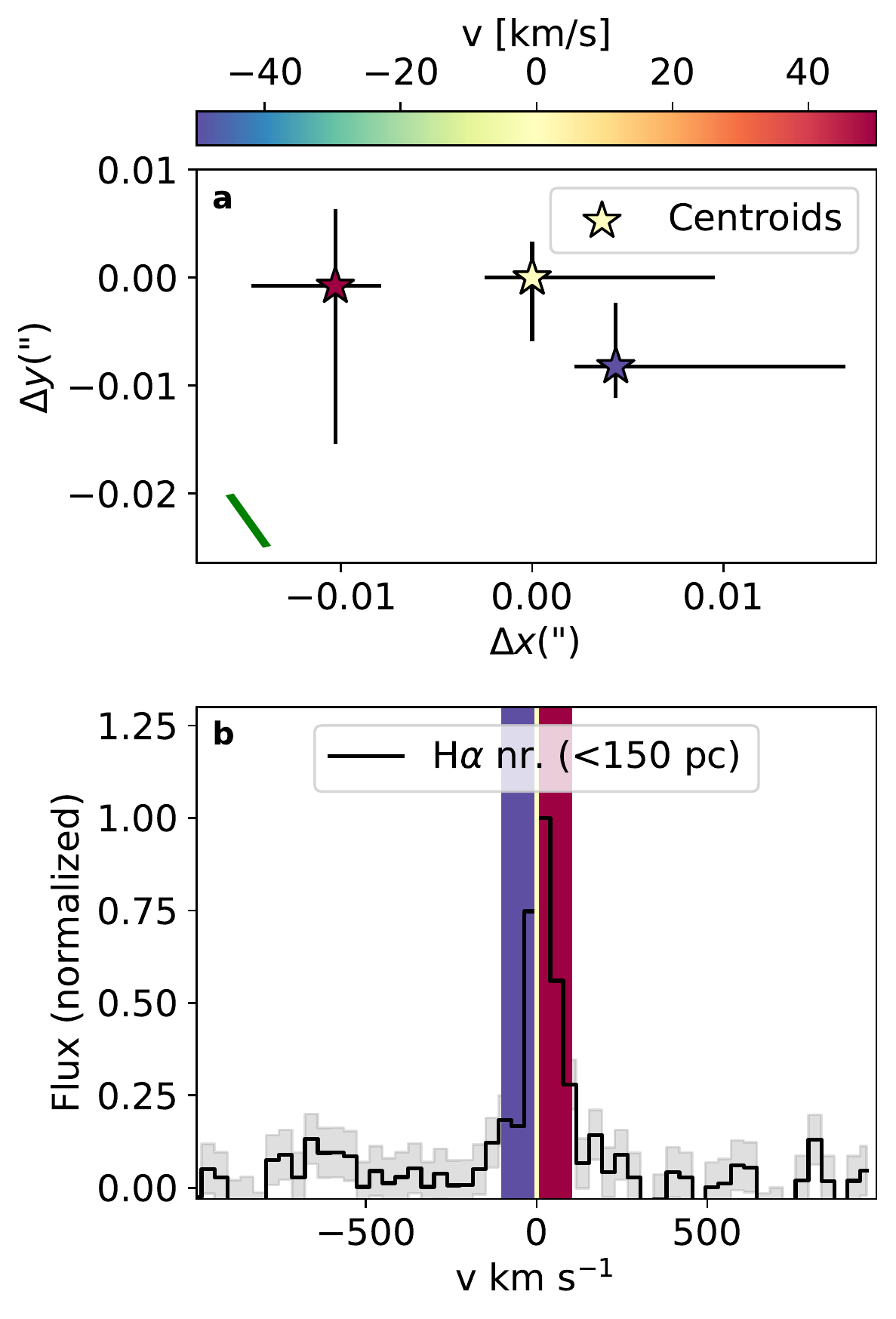}
    \caption{\textbf{Spectroastrometry on the narrow line.} Panel \textbf{a} - locations of the centroids of each velocity channel with the green line showing the PA of the IFU slicer. Panel \textbf{b} narrow \Halpha line profile extracted from a 0.1" aperture (coinciding with the scales traced by our kinematics fits) with colored bars showing the velocity bins. While the line profile appears to show some evidence of a wing at $\sim -150$~km~s$^{-1}$, expanding the velocity channels to include it does not meaningfully alter the results.}
    \label{fig:spectro_centroids}
\end{figure}

We perform the spectroastrometric analysis using the same broad line and continuum subtracted cube as in the previous section. The centroid measurements were carried out using three velocity channels: $-40 < v < 40$~km~s$^{-1}$, serving as the zero velocity reference point and $-100 < v < 0$~km~s$^{-1}$, $0 < v < 100$~km~s$^{-1}$, tracing the blue and red wings of the line respectively. The widths of these channels are equal to 2-3 spectral channels of the underlying cube and were chosen in order to maximize S/N in each channel image. The centroid positions were measured by fitting 2D quadratic polynomials (although the results are not strongly influeced by the exact centroiding routine) via \texttt{astropy}, with the errors estimated through bootstrap resampling utilizing the rms of the blank parts of the image as 1$\sigma$ error on the flux. The centroid positions are shown in Extended Data Fig.\ref{fig:spectro_centroids}a and indicate that the centroids of the blue and red wings of the line lie on the opposite ends of the $v = 0$ channel centroid and are broadly aligned with the large scale rotation. While the exact positions of the centroids are subject to some uncertainty, the significance of the separation is 3-5$\sigma$, indicating that the measurements trace small-scale rotation. It should be noted that the $|v| = 0$ centroid in Extended Data Fig.\ref{fig:spectro_centroids}a does not coincide with the dynamical center inferred by \texttt{MOKA3D} modeling, this is because the overall flux distribution (Fig. \ref{fig:rot_curve}a) is asymmetric. \texttt{MOKA3D} takes into account this asymmetry (as well as any other non-radial flux distribution), however, spectroastrometry does not. As a consequence, the absolute location of the spectro-astrometry centroids is affected by this issue, but the relative position of the centroids is very precise.

The integrated narrow \Halpha profile is shown in Extended Data Fig. \ref{fig:spectro_centroids}b and indicates some emission present at $|v| \sim 150$~km~s$^{-1}$ at about 2$\sigma$ significance, however, extending the velocity bins out to 150~km~s$^{-1}$ does not impact the results.

In order to estimate the virial mass of the object, we utilize the methods laid out in \citem{Gnerucci2011}, which derive the following mass estimator:
\begin{equation}
\label{eq:vir}
    M_{\rm spec}\sin^2{i} = f_{\rm spec}\frac{\mathrm{FWHM}^2r_{\rm spec}}{G},
\end{equation}
where $r_{\rm spec}$ is the spectroastrometric radius, FWHM - width of the emission line analyzed and $i$ - the inclination angle. The calibrated $f_{\rm spec}$ value is 1.0, with 0.15~dex scatter \citem{Gnerucci2011}.

The measured separation between the red and blue centroids of the H$\alpha$ is $24.9_{-9.4}^{+9.4}$~pc, after correcting for the shear lensing factor of 3.52 \citem{Furtak2023_AGN}. Taking $r_{\rm spec}$ as half this separation gives $r_{\rm spec} = 12.4_{-4.7}^{+4.7}$~pc, coupled with $\mathrm{FWHM}=52_{-14}^{+14}$~km~s$^{-1}$ \citem{DEugenio2025}, this yields $\log{M_{\rm spec}\sin^2{i}/\mathrm{M}_{\odot}} = 6.90_{-0.23}^{+0.23}$. This estimate should be interpreted as a lower limit as the inclination is unconstrained.

\subsection*{Rotation curve analysis}
In order to better constrain the central point mass as well as investigate the alternative scenarios, such as potential contributions from a compact nuclear star cluster, we combine our spectroastrometric measurements with resolved kinematics in order to construct a rotation curve for QSO1.

Computing the velocity field to which models can be fit requires combining line of sight velocity ($v_{\rm obst}$, shown in Fig. \ref{fig:rot_curve}b) with observed velocity dispersion to obtain the corrected, 'rotational' velocity $v_{\rm rot}$. In order to do this, we add the velocity dispersion and $v_{\rm obs}$ in quadrature. The dispersion value was 15-20~km~s$^{-1}$ for the red and blue wings and $\sim34$~km~s$^{-1}$ for the axis of rotation. In order to compute $v_{\rm rot}$ for the spectroastrometric data points, we use a flux weighted average of the velocity channels, giving $v_{\rm obs}\sin{i} = 51$~km~s$^{-1}$. Hence, the dispersion corrected $ \langle v_{\rm rot}\sin{i} \rangle = 20 \pm 6$~km~s$^{-1}$ in the extended wings and $v_{\rm rot}\sin{i} = 61 \pm 6$~km~s$^{-1}$ for the spectroastrometric points, indicating a better 3$\sigma$ detection of rotation. The factor of $\sin{i}$ is written to explicitly state that these are the projected values, uncorrected for inclination.

As the resultant rotation curve, shown in Fig.\ref{fig:BHmeas}, is sparsely sampled, we only consider simple 1 to 2 parameter models for fitting. Model curves were constructed following a Keplerian prescription:
\begin{equation}
\label{eq:rot_curve}
    v(R) = \sqrt{\frac{GM(<R)}{R}}
\end{equation}
where $R$ is the distance from the center, while $M(<R)$ - the mass enclosed within $R$. Our fiducial fit follows a point mass assumption with $M(<R) \equiv M \equiv const.$, which yields $\log{M/M_{\odot}} = 6.94 \pm 0.05$. We note that the uncertainty on this value is purely a fitting error and could be underestimated, hence we re-estimate the error using bootstrap resampling, taking into account the width of the velocity bins, we estimate a more conservative measurement error of 0.15~dex.

\begin{figure}
    \centering
    \includegraphics[width=\linewidth]{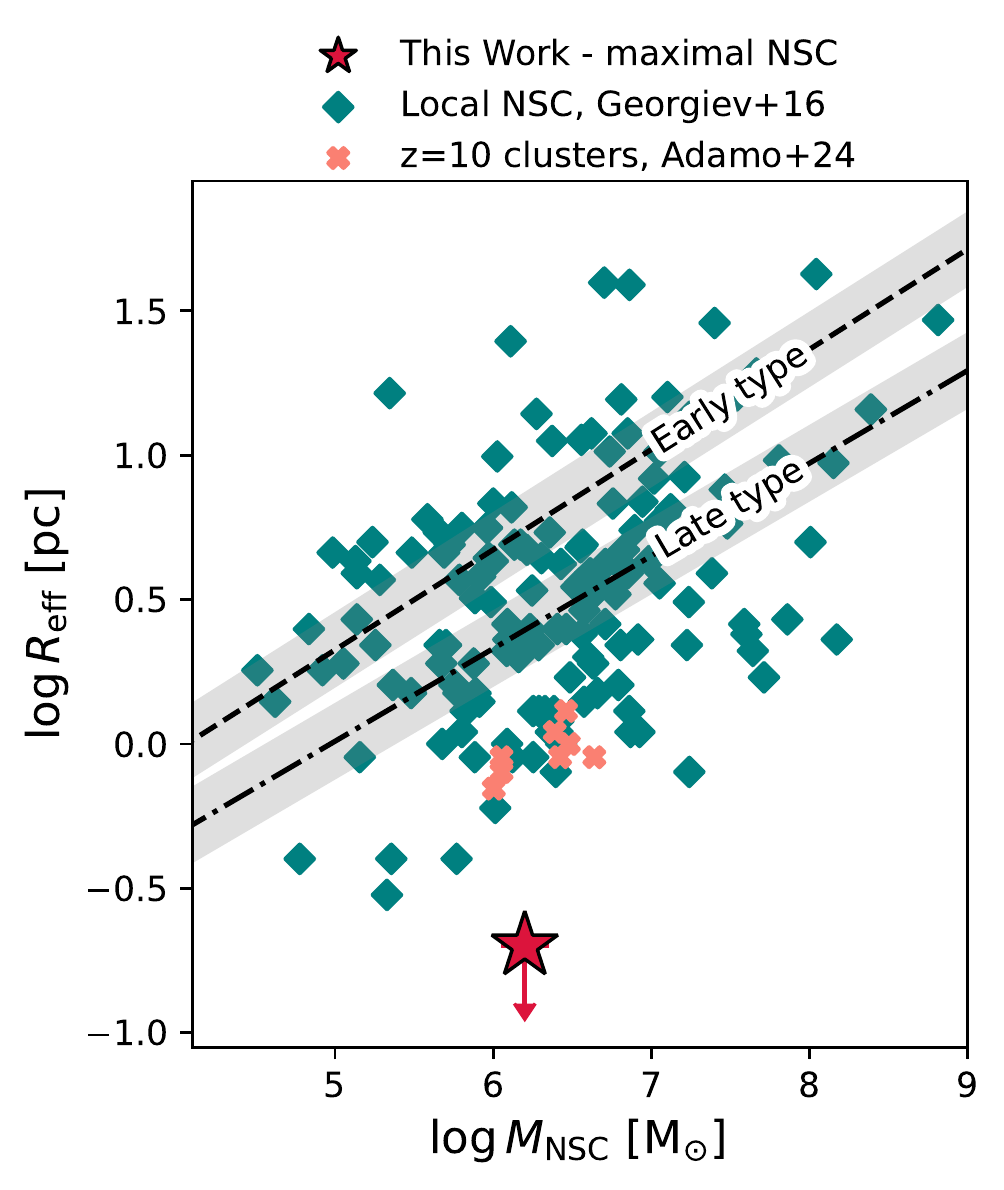}
    \caption{\textbf{Comparison between QSO1 and star cluster observations on the mass-radius plane.} The upper limit on the size of a NSC in QSO1  is shown by a red star, illustrating that it is implausible when compared with local  NSC (teal diamonds) \protect\citemet{Georgiev2016} and dense star clusters at z=10 (salmon crosses) \protect\citemet{Adamo2024}. The black lines illustrate mass - radius scaling relations for Early and Late type galaxies derived by Ref. \protect\citemet{Georgiev2016} with grey shading indicating scatter. The value for QSO1 represent the maximal extent allowed by an NSC-only model. The upper limit on the radius is nearly an order of magnitude lower than both local and distant clusters at a similar mass range.}
    \label{fig:nsc_max_extent}
\end{figure}

In order to fit the curve with a compact stellar mass distribution, we fit the data with a nuclear star cluster (NSC) model derived for the Milky Way by \citem{Schodel2014} who find a density profile following an $R^{-2}$ power law in the central 5~pc and dropping off as $R^{-3}$. From this density profile, we construct the following function for $M(<R)$:
\begin{equation}
    M(<R) = \begin{cases}
    4\pi RA & \text{if } R<R_c\\
    4\pi R_cA\left[1 + \log{\left(\frac{R}{R_c}\right)}\right] & \text{if } R \geq R_c
    \end{cases}
\end{equation}
where  $A$ is the parameter setting the overall normalization and $R_c$ is the radius at which the switch in the power law profile occurs. It is important to note that this is different from the ``effective radius'' $R_e$
of the 2D light distribution; in the case of the MW NSC the $R_e$ is a factor of 0.84 smaller than $R_c$.   We initially fit the NSC model fixing $R_c$ to 5~pc, the same value as found by Ref.\citem{Schodel2014}. This fit is shown as a dashed gray line in Fig.\ref{fig:BHmeas} and produces a considerably worse $\chi^2_{R} = 3.8$ than the point-mass (pure BH) fit with $\chi^2_{R} = 0.8$. This corresponds to a difference in the BIC of $\sim 20$, indicating robust preference for the point mass fit. It $R_c$ is allowed to vary freely, then the NSC best fit model gives $R_c = 10^{-4}$~pc with $M(<R_c) \sim 10^6$~M$_{\odot}$, which would imply extreme stellar densities, in excess of $10^{17}$~M$_{\odot}$~pc$^{-3}$. Such densities are orders of magnitude above the densest stellar systems seen in the Universe and show that our NSC model effectively collapses to a point mass if $R_c$ is not fixed. We estimate an upper limit on $R_c$ by fitting a fixed value and lowering it until the difference in BIC between the best-fit and the fixed $R_c$ model reduces below 5. This way we estimate $R_c < 0.2$~pc with $M(<R_c) \sim 10^{6.2}$~M$_{\odot}$; this limit is over 1~dex below even the most compact NSC in this mass range in the local Universe \citemet{Georgiev2016}, as well as the dense star clusters found in the lensed Cosmic Gems arc by Ref. \citemet{Adamo2024}, as illustrated in Extended Data Fig. \ref{fig:nsc_max_extent}. The upper limit on the NSC stellar mass is even lower if one adopts the density profile inferred for NSCs in other galaxies ($\rho \propto r^{-2}$)\citemet{Pechetti2020}. Hence, a point mass is needed to account for our observed dynamics.

In addition to the NSC model described above we consider the Plummer sphere \citem{Plummer1911} model, frequently used to describe the density profiles of globular clusters. The enclosed mass function for the Plummer sphere takes the following form:
\begin{equation}
    M(<R) = M_0\frac{R^3}{\left(R^2 + a^2\right)^{3/2}}
\end{equation}
where $M_0$ is the total mass of the system and $a$ - the scale radius. While this model performs similarly well to the point mass, it does so by fitting $a \sim 10^{-4}$~pc and thus results in similar unphysically high stellar densities, as in the case of the NSC.

In addition to being excluded by the $\Delta \mathrm{BIC}$ value, a Milky Way NSC-like density profile with $R_c \sim 5$~pc produces a total stellar mass of $10^{7.2}$~M$_{\odot}$, which is above constraints on the stellar mass derived from UV and optical emission as discussed in the next section. 

A remaining potential caveat of our analysis is that only isotropic velocity distributions were considered - velocity anisotropies could steepen the radial velocity gradient of a diffuse mass component \citemet{Gebhardt2005} increasing the allowable extended mass. However, the steepness of the observed velocity gradient is such that any extended component collapses to a point when the scale radius is left free. Hence, it is unlikely that anisotropies of the underlying velocity field significantly skew our results.

Lastly, we explore what, if any, constraints on the dark matter (DM) halo surrounding the object can be obtained from our data. We thus fit the widely adopted Navarro-Frenk-White density profile \citemet{NFW_halo}. The enclosed mass for which is given by:
\begin{equation}
    M(<R) = 4\pi\rho_0R_s^3\left[\ln(\frac{R+R_s}{R_s}) - \frac{R}{R+R_s}\right],
\end{equation}
where $\rho_0$ and $R_s$ are the characteristic density and scale radius respectively. However, as with the previous extended mass distributions, the above model collapses to a point with $R_s \sim 10^{-4}$ and $\rho_0 \sim 10^{14}$~M$_{\odot}$~pc$^{-3}$, once more collapsing to a point mass and producing unrealistic densities. However, this does not imply that QSO1 resides outside of a DM halo. Instead, our attempts at reproducing the kinematics with an extended density profile imply that any extended mass component is sub-dominant at the $<200$~pc scales probed by our measurements. 

\subsection*{Optical and UV constraints on stellar mass}
The dynamical mass estimates discussed above are sensitive to the total mass present in the object. The kinematics fits shown in Fig.\ref{fig:BHmeas} alone constitute $>5\sigma$ evidence against significant stellar contribution to the mass of the object. In addition, Ref.\citem{Ji2025} have shown that the optical continuum of Abel2700-QSO1 is dominated by AGN emission, with a stellar contribution of $<$10\%. Stellar contribution to the UV continuum can not be ruled out and was indeed suggested as a way to explain the observed power law slopes by Ref.\citem{Naidu2025}, and the observed spatial offsets \citemet{Torralba2025}, although some indications of UV variability in QSO1\citem{Ji2025} suggest that even the UV maybe AGN-dominated. Yet, even conservatively assuming that the UV light is entirely stellar, it would give a stellar mass much lower than the dynamical mass inferred by us, as discussed in the following. The UV magnitude measured for image A is $M_{\rm UV} = -15.60$, after correction of $\mu=6.2$. Utilizing the $M_* - M_{\rm UV}$ relation from Ref.\citemet{Simmonds2024}, we obtain a stellar mass estimate of $\log{M_*/\mathrm{M}_{\odot}} = 6.03$. This is almost 1~dex lower than the measured virial mass. In addition, while Ref.\citem{Ji2025} have shown that the UV continuum is damped by a different hydrogen column density than the optical, the observed slope can be explained by nebular emission. Hence our stellar mass value should be treated as an upper limit and stars can be conclusively ruled out as dominant contributors to the virial mass budget.

\subsection*{MOKA3D kinematics modeling}

The measurements described above, while self consistent, do not fully take into account instrumental effects such as PSF broadening. In order to check the robustness of our conclusions if such effects are accounted for, we refit the narrow line cube with the \texttt{MOKA3D} framework \citem{Marconcini2023, Marconcini2025}. \texttt{MOKA3D} is a 3D kinematic framework that can model conical outflows or disks by assuming spherical, conical or cylindrical geometries respectively, with any irregular distribution of the emitting clouds within the velocity field. The 3D model is populated with a distribution of fictitious clouds that account for the observed emission. These clouds are weighted according to the observed narrow H$\alpha$ flux in each spaxel and spectral channel of the data cube. As described in the following, the model clouds follow an analytical velocity field as a function of the radius, whose parameters are fitted to reproduce the observed emission and kinematic features via least-squares minimization.

\begin{figure*}
    \centering
    \includegraphics[width=\linewidth]{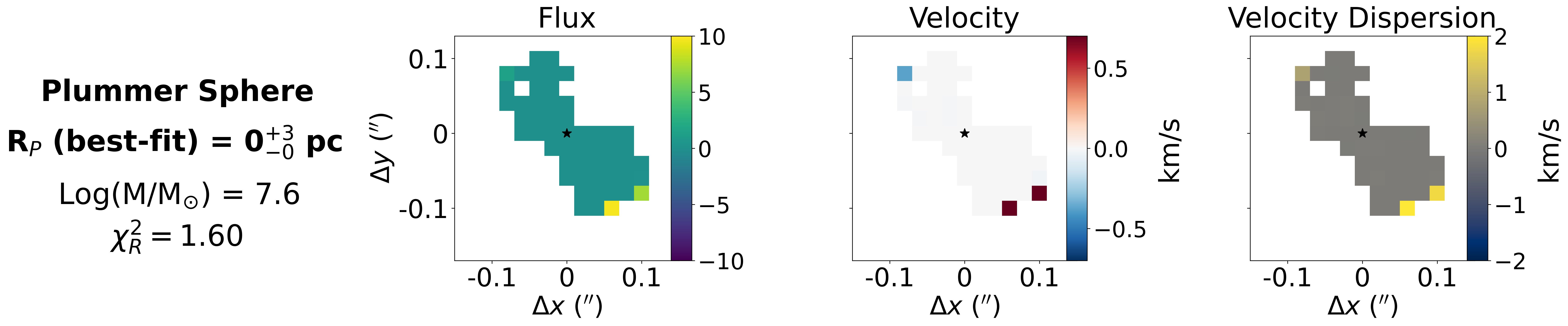}
    \caption{\textbf{Residuals of the \texttt{MOKA3D} fit of the Plummer sphere model.} While the performance of this model is effectively only slightly worse than that of a point mass ($\chi^2_R=1.6$ vs $\chi^2_R=1.17$), the best fit $R_p = 0^{+3}_{-0}$~pc indicates that this is simply because the model ends up reproducing a point mass.}
    \label{fig:MOKA3D_plummer}
\end{figure*}

As in the previous analysis, we consider three potential mass distributions - a point mass, a NSC density profile  and a Plummer sphere. The comparison between the residuals of the different models is given in Fig. \ref{fig:BHmeas} and Extended Data Fig. \ref{fig:MOKA3D_plummer}. As can be seen in the figures, a point mass profile is strongly preferred over any extended density profile. The Plummer sphere model, while leaving similar residuals to a Keplerian curve, does so with a scale radius $R_{P} = 0^{+3}_{-0}$~pc (Extended Data Fig. \ref{fig:MOKA3D_plummer}), essentially collapsing into a BH. The NSC model remains extended even when $R_c$ is allowed to vary (best fit $R_c = 4 \pm 2$~pc), however, it leaves significant systematic residuals in both velocity and velocity dispersion profiles, as shown in Fig. \ref{fig:BHmeas}. While the absolute values of the velocity residuals are 0.5 - 1.0~km~s$^{-1}$, their systematic nature (and high $\chi^2_R$) indicates that the NSC model is inadequate in modeling the observed kinematics.

It is notable that the \texttt{MOKA3D} point mass model gives $\log{M_{\rm BH}/M_{\odot}} = 7.7 \pm 0.3$ with an inclination of $52^\circ \pm 2^\circ$, entirely consistent with the lower limit obtained from the previous direct measurements. In fact, the spectroastrometric mass estimate, when corrected for the inclination becomes consistent with the \texttt{MOKA3D} value to within 2$\sigma$ ($7.2 \pm 0.15$ vs $7.7\pm 0.3$).

\begin{figure}
    \centering
    \includegraphics[width=\linewidth]{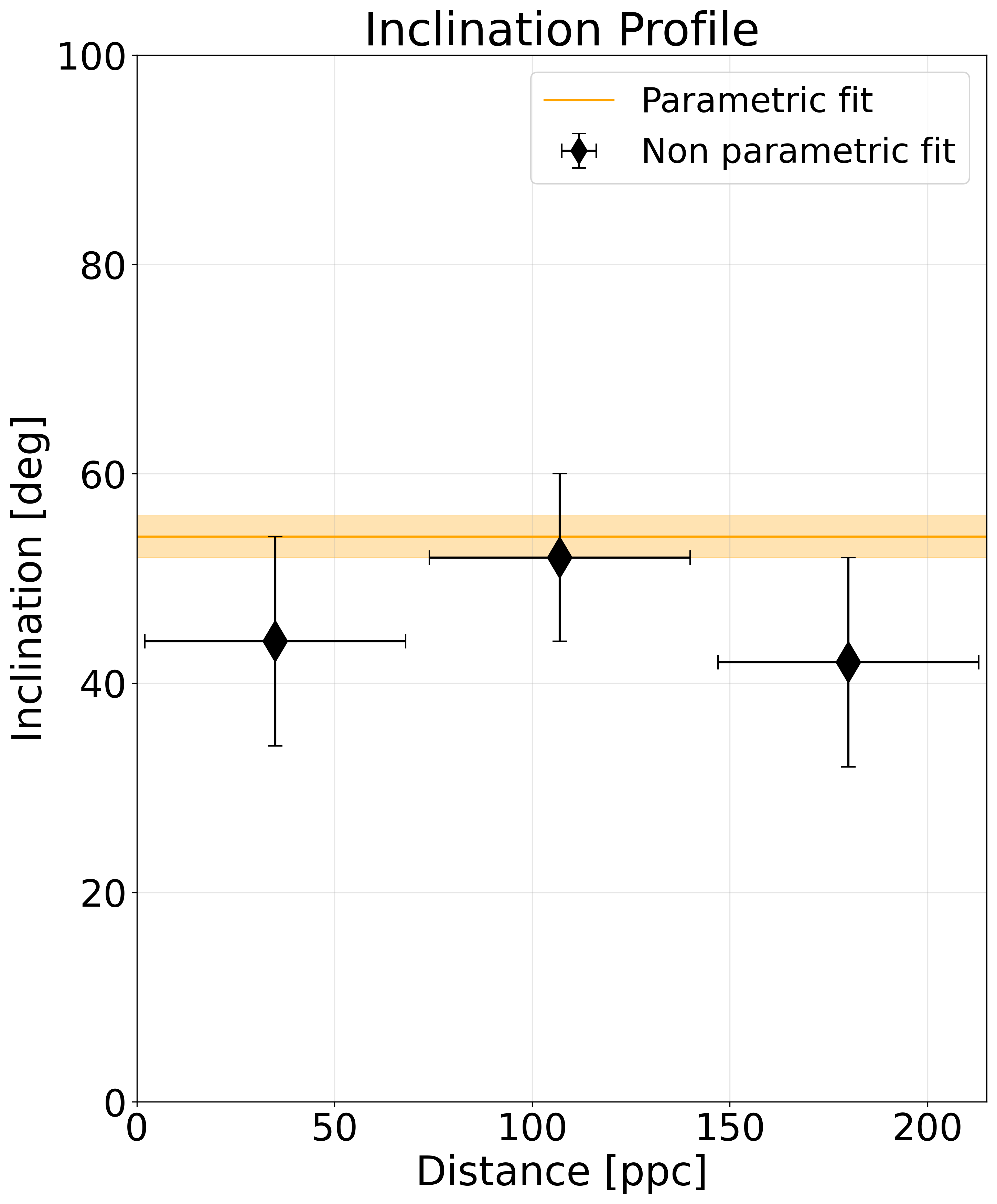}
    \caption{\textbf{\texttt{MOKA3D} inclination constraints.} Inclination value found by fitting parametric models, orange line with shading indicating 1$\sigma$ uncertainty, compared to inclinations found from a non parametric disk model in three independent rings (black points). The non parametric model is internally consistent among the three independent rings and consistent with the parametric curves well within 1$\sigma$.}
    \label{fig:inclination}
\end{figure}

In order to test if the inclination estimate is robust, as well as verify the presence of a rotating disk, we construct a non-parametric model wherein the disk is split up into three distinct shells that are fitted with independent inclinations. We find that this model fits the data well and produces shell inclinations of $\sim 45 \pm 10^\circ$ consistent with each other and the value found by the parametric models (Extended Data Fig. \ref{fig:inclination}). The fact that consistent results are obtained regardless of the precise analytical procedure employed indicates that our measurements are robust.

We finally emphasize that the MOKA3D analysis is independent of the spectro-astrometry approach.

\subsection*{Tentative nuclear outflow}

The remaining concern of our kinematics modeling is the potential presence of outflows, which are common in AGN host galaxies in the local Universe and a biconical outflow viewed edge-on can resemble a rotating disk. This is a particular concern for QSO1 as analysis of the H$\beta$ and [OIII] emission in this source by\citem{Maiolino2025} has identified a likely extended outflow component in H$\beta$, traced by an intermediate width, blueshifted component. In addition, the broad H$\alpha$ line shows a bimodal profile, with a narrower (FWHM = 490~km~s$^{-1}$) broad component superimposed on a broader (FWHM = 1800~km~s$^{-1}$) emission line \citem{DEugenio2025}.

\begin{figure}
    \centering
    \includegraphics[width=\linewidth]{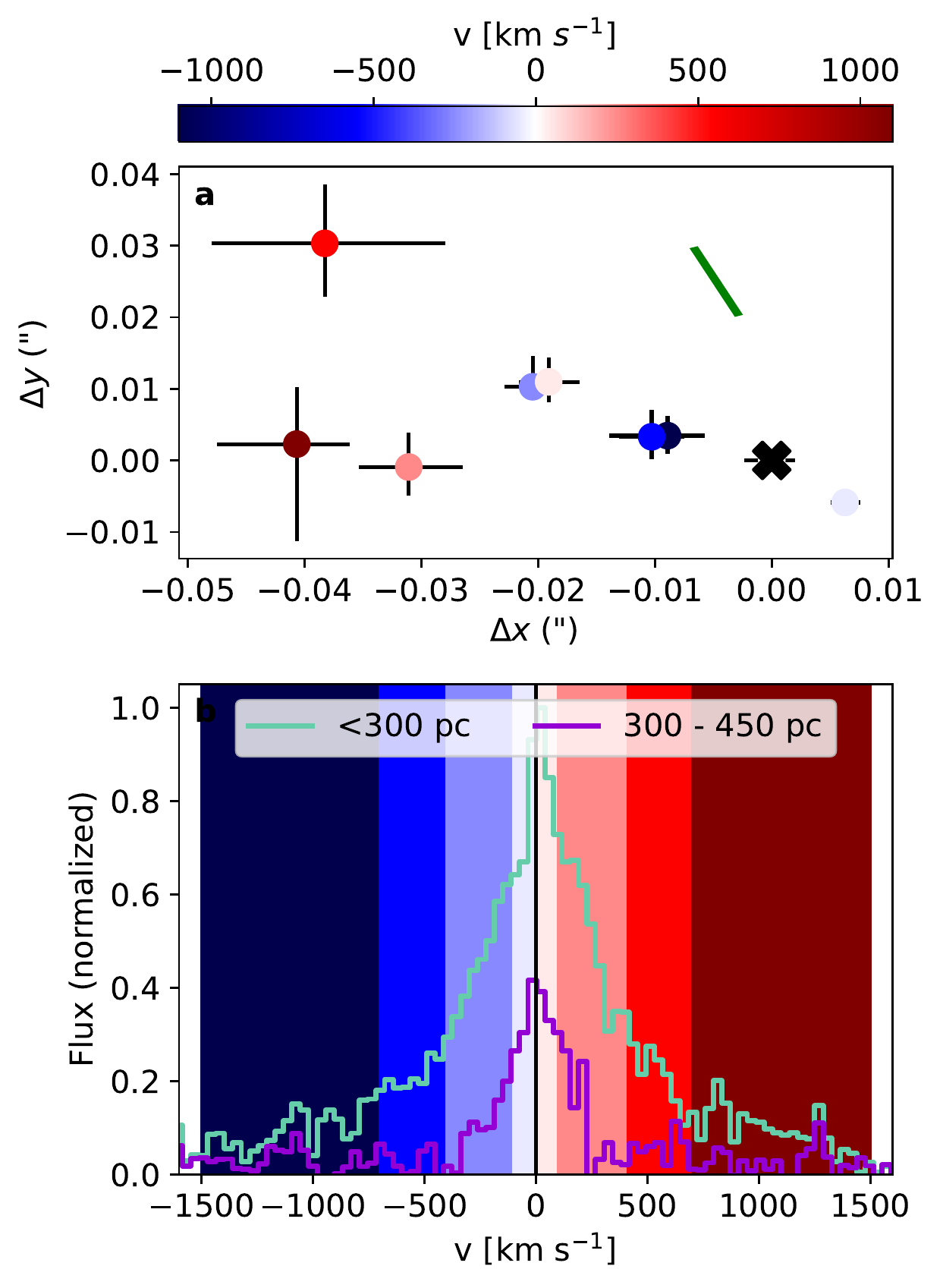}
    \caption{\textbf{Spectroastrometric analysis of the full line profile.} Panel \textbf{a} Centroids of the different velocity channels defined in the text showcasing the offset of the positive high velocity channels. The $|v| = 0$ channel is marked with a black cross while the green line shows the PA of the IFU slicer. Panel \textbf{b} \Halpha line profiles extracted from the inner 300~pc (green line) and an 300 - 450~pc annulus (violet line) normalized to the peak of the inner 300~pc spectrum. The vertical colored bars denote the velocity bins for which centroids were measured with colors corresponding to their points in \textbf{a}.}
    \label{fig:outflow_spectroastro}
\end{figure}

We investigate weather the intermediate \Halpha component is spatially distinct from the broad one by extracting an annular spectrum between 0.2 and 0.3" from the center - corresponding to 300 - 450~pc on the source plane. This extraction annulus is $\sim 30 \%$ larger than utilized in \Hbeta analysis by Ref. \citem{Maiolino2025} in order to account for additional beam smearing, by a 30\% larger PSF at \Halpha wavelengths. The resulting annular spectrum is shown in Extended Data \ref{fig:outflow_spectroastro}b (purple line) and clearly reveals that the intermediate velocity component (|v| < 500~km~s$^{-1}$) is spatially distinct from the broad component and is thus likely associated with a large scale outflow.

In order to further investigate the presence of outflows in our object, we perform spectroastrometric analysis of the full line profile, splitting it into channels of $|v| = 50$~km~s$^{-1}$, tracing low velocity gas, $|v| = 250$~km~s$^{-1}$ and $|v| = 550$~km~s$^{-1}$ tracing the intermediate component and $|v| = 1100$~km~s$^{-1}$ - tracing the high velocity broad wings. The centroids of the relevant velocity channels are shown in Extended Data Fig.\ref{fig:outflow_spectroastro}a and show that the red wing of the line is significantly offset from the blue. This likely indicates contamination by higher velocity unresolved outflows inducing turbulence as the red and blue nodes of the high velocity channels do not lie on opposite sides of the $|v| = 0$~km~s$^{-1}$ channel. In addition, the apparent 'rotation' is more perpendicular to the IFU slices, which might be indicative of  instrumental effects associated with the calibration of different slices. 

While all broad components are removed by our broad line subtraction procedure, the  narrow line kinematics may still end up contaminated. Indeed, the centroid of the red wing narrow line slightly shifts in position once the underlying broad emission is subtracted (see Extended Data Fig. \ref{fig:spectro_centroids}), although the significance of this shift is $<2\sigma$. However, as shown in Fig. \ref{fig:rot_curve}, the observed velocity dispersion of the narrow component is very low and drops off with increasing distance from the axis of rotation, which would be unusual for an extended outflow. In addition, it would be a very improbable conspiracy for the velocity distribution of the putative outflow to exactly mimic Keplerian rotation.

\begin{figure*}
    \centering
    \includegraphics[width=\linewidth]{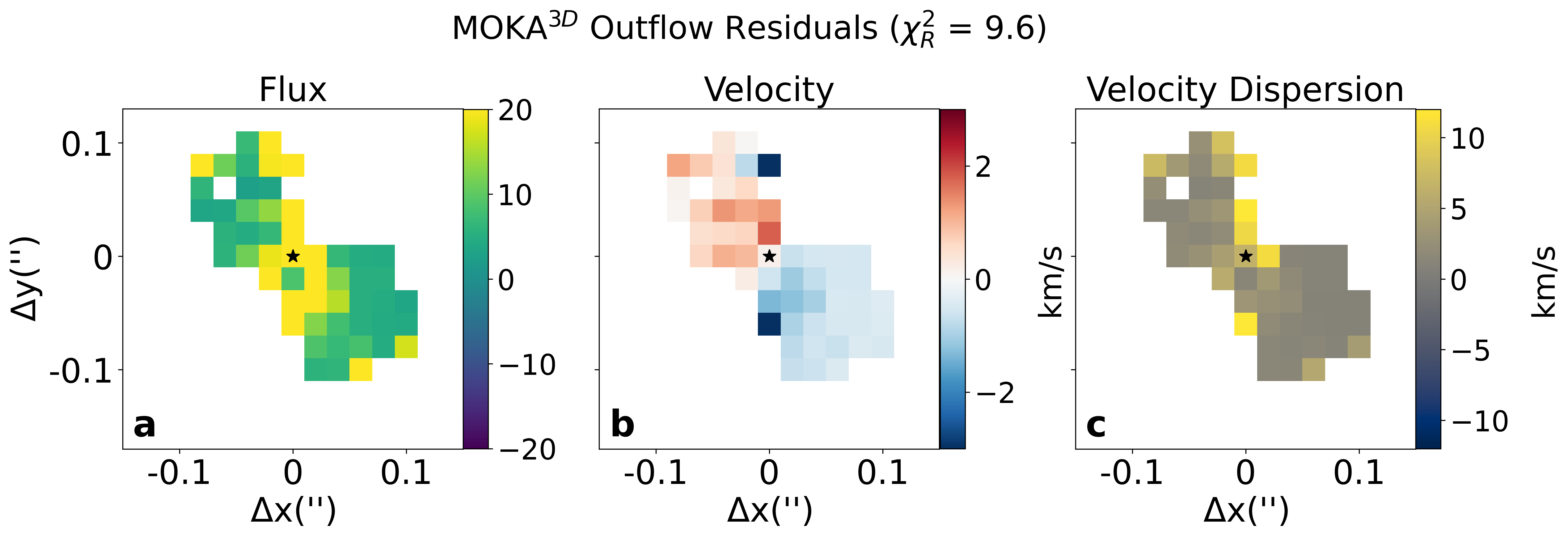}
    \caption{\textbf{Residuals of a fit assuming a biconical outflow.} The residuals and hence the $\chi^2_R$ value are considerably worse than any of the fits assuming a rotating disk.}
    \label{fig:outflow_res}
\end{figure*}

To  investigate more quantitatively the outflow potential contamination to narrow line kinematics we construct a biconical outflow model in \texttt{MOKA3D}. In particular, we assume a bi-conical geometry with aperture angle of 40$^{\circ}$, position angle of 135$^{\circ}$\footnote{The position angle is measured clockwise from the north.} and a constant radial velocity field for both the approaching and receding cones. We constrained the inclination angle of the blue-shifted cone to be in the range [-90$^{\circ}$,90$^{\circ}$], i.e. we assumed that the blue-shifted cone is approaching the observer. As shown in Extended Data Fig.\ref{fig:outflow_res}, we found that the best solution to reproduce the observed kinematic features requires an outflow intrinsic velocity of 110~km~s$^{-1}$ and inclination with respect to the line of sight of 85$^{\circ}$, i.e. only 5$^{\circ}$ inclination with respect to the plane of the sky.
As shown in Extended Data Fig. \ref{fig:outflow_res} the biconical outflow model still leaves systematic residuals resembling a rotating disk resulting in $\chi^2_R = 9.6$, considerably worse than any of the models assuming a rotating disk. This, coupled with the fact that the measured velocity dispersion drops considerably away from the axis of rotation (Fig. \ref{fig:rot_curve}) indicates that, while outflows are likely present in this source, the narrow line kinematics are relatively undisturbed by them.

\subsection*{BH mass estimate assuming electron scattering}
In order to test how the BH mass would change if the dominant broadening mechanism was electron scattering, as proposed by \citem{Rusakov2025}, we refit the H$\alpha$ and H$\beta$ lines of QSO1 with a scattering profile constructed via the following equation:
\begin{equation}
    P(\lambda) = f_\text{scatt} (E*G_\text{BLR})(\lambda) + (1 - f_\text{scatt}) G_\text{BLR}(\lambda),
\end{equation}
where $G_\text{BLR}(\lambda)$ is the intrinsic broad-line profile, assumed to be virially broadened and Gaussian in shape, $f_\text{scatt}$ - the fraction of light affected by the electron scattering and $E(\lambda)$ - the exponential electron scattering profile \citemet{Laor2006}, defined as
\begin{equation}
    E(\lambda) \propto e^{- \frac{|\lambda - \lambda_0 |}{W}},
\end{equation}
where $\lambda_0$ is the central wavelength, and $W$ is the width of the exponential.

\begin{figure}
    \centering
    \includegraphics[width=\linewidth]{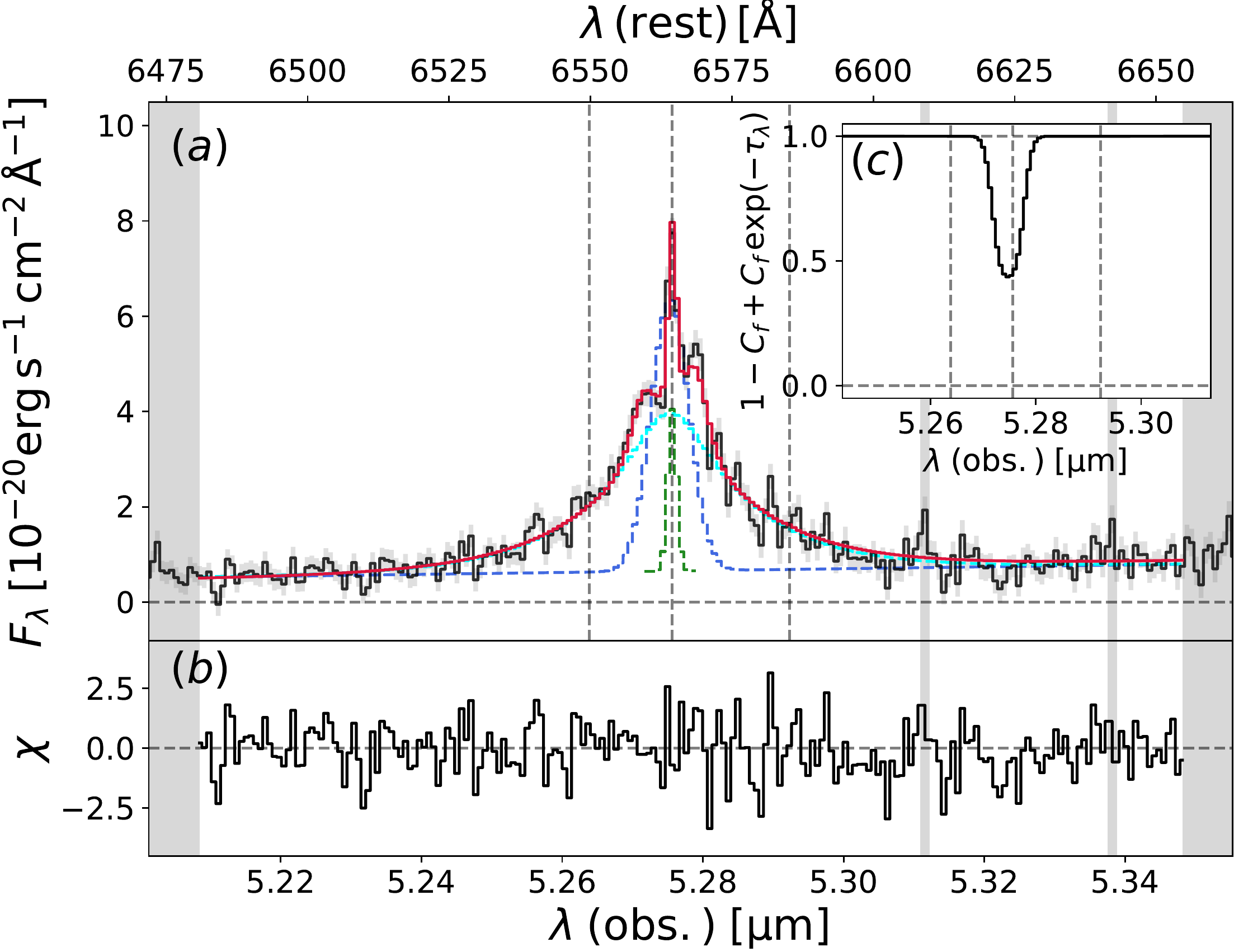}
    \caption{\textbf{Fits to the \Halpha line assuming a line profile dominated by electron scattering.} The combined best fit is shown by the red line, the scattered \Halpha BLR is shown in cyan with the NLR shown in green, the blue line shows the transmitted, unscattered BLR. As can be seen in residual plots (panel b), the model fits the data well, however, the BH mass derived is 2 ~dex ($\sim4\sigma$) below the dynamic mass measurements as discussed in the text.}
    \label{fig:scattering}
\end{figure}

The absorption features in the Balmer lines were modeled using \autoref{eq:abs}, as in the previous section. The priors on the fit were uninformative, flat priors over all model parameters. We inspected the posterior probability over each model parameter, and verified that no posterior hit the boundaries (except for physically motivated boundaries, such as non-negative emission-line flux). The posteriors were estimated using the Monte Carlo sampler \texttt{emcee} \citemet{emcee}, using ten walkers per free parameters, and initializing
the chains inside a sphere centered on the maximum-likelihood solution, obtained from
differential evolution \citemet{storn+price1997}. The best-fit, produced by the medians of the posteriors, is shown in Extended Data Fig. \ref{fig:scattering}. As can be seen there, the integrated spectrum is well fitted by this model producing a BH mass estimate of $\log{M_{\rm BH}/M_\odot} = 5.9 \pm 0.1$. We note that, as discussed in the main text and shown in Fig. \ref{fig:mbh_comp}, this value is nearly 2~dex lower than the direct mass estimates, showing that it is implausible that the observed \Halpha line profile can be attributed to electron scattering. In case of the BH mass being this low, extreme stellar densities would be required to produce a point mass consistent with the one measured. This result confirms the findings of \citemet{Brazzini2025}: despite providing
an excellent fit to the data, the electron scattering model violates basic physical constraints.

\subsection*{Dynamic upper limit on the stellar mass in the host galaxy}
In addition to spectroscopic constraints on the stellar mass of the host, we endeavor to estimate the maximum extended mass component permitted by our data. We perform this estimate by constructing a combined point mass and an extended exponential disc component modelled as an $n = 1$ S\'ersic density profile of half mass radius 100~pc, truncated at 300~pc - the maximum extent of the source found by \citem{Maiolino2025}. We note that galaxies with such a small stellar mass at such high redshift are observed to be on average smaller than this radius\citemet{Miller2025sizes}.

If both components are left free, the point mass dominates, essentially recovering the previous pure Keplerian fits. We thus derive an upper limit on the extended mass component by forcing a particular extended mass value and increasing it until the $\Delta BIC > 5$ between the fiducial and extended component models. Following this method, we derive the maximum extended mass of $\log{M\sin^2{i}} = 6.85$, similar to the mass of the BH.

However, this extended mass component is a composite of contributions from stars, gas and Dark Matter and thus only serves as an upper limit to the total stellar mass. We tighten this limit further by estimating the ionized gas mass from the narrow line emission. The ionized gas mass can be estimated from narrow \Hbeta emission via the following equation:
\begin{equation}
    \label{eq:mion}
    M_{\rm ion} = 0.8\frac{m_{\rm p} \, L_{\Hbeta}}{j_{\Hbeta}\,n_{\rm e}},
\end{equation}
where $L_{\Hbeta}$ is the \Hbeta line luminosity, $j_{\Hbeta}$ - \Hbeta emissivity, $n_e$ - electron density and $m_p$ - proton mass. Assuming standard warm interstellar medium conditions of $T \sim 10000$~K and $n_e \sim 100$~cm$^{-3}$ and taking $L_{H\beta} =  5\times 10^{40}$~erg~s$^{-1}$ from \citem{Ji2025} we obtain $\log{M_{ion}/M_{\odot}} = 6.7$, with a similar mass obtained by using narrow \Halpha measurements from \citem{DEugenio2025}. Utilizing the inclination of $52 \pm 2^{\circ}$ derived from \texttt{MOKA3D}, we estimate that the effective contribution of ionized gas to the uncorrected kinematics is $\log{M_{ion}\sin^2{i}} = 6.5$. Hence, the dynamical contribution of an extended stellar component cannot exceed $\log{M\sin^2{i} - M_{ion}\sin^2{i}} = 6.6$.

In addition to an extended stellar distribution we also consider the maximal contribution of a compact NSC. For this, we fix $R_c$ to 0.5~pc, a factor of 2 below the most compact NSCs seen in the local Universe \citemet{Georgiev2016} to account for higher redshift structures being less relaxed. As with the previous attempts to fit a non-point mass distribution, the mass of the extended component goes to zero if left free. We thus estimate an upper limit on the NSC mass through requiring $\Delta BIC <5$ from the fiducial model. The upper limit on the NSC mass is thus $M(<R_c)\sin^2{i} = 10^{6.1}$~M$_{\odot}$ with the BH mass reducing slightly to $M_{\rm BH}\sin^2{i} = 10^{6.44}$~M$_{\odot}$. A similar mass ratio is maintained even when setting $R_c = 0.2$~pc, the upper limit found earlier.

We also repeat the above analysis using the \texttt{MOKA3D} framework, finding that $M_{*} \leq 10^{7.2}$~M$_{\odot}$ within $\sim$800~pc (i.e. well beyond any possible galaxy extension at this redshift)\citemet{Miller2025sizes} for an extended exponential disc profile and $M_{*} \leq 10^{6.55}$~M$_{\odot}$ for a NSC-like distribution with $R_c = 0.5$~pc. Taking the estimated $M_{\rm BH} = 10^{7.7}$~M$_{\odot}$, we obtain $M_{BH}/M_{star} > 2 - 15$. As before, the stellar masses estimated above are  upper limits, as dynamical mass measures are sensitive to the combined mass of gas, stars and DM, which we are unable to fully disentangle due to lack of emission tracing gas on larger scales. We note that these dynamical upper limits are fully consistent with the gas and stellar masses estimated in the previous section, overall, any extended mass component below $10^{6.5}$~M$_{\odot}$ has no impact on the fit given the observational uncertainties.

The above analysis implies a $M_{\rm BH}/M_{star} > 2$ for any reasonable mass distribution, with the inclination term canceling in the ratio. This is an extreme lower limit since it does not include contributions from non-ionized gas and DM. Therefore, QSO1 represents the most naked BH yet discovered, leaving only direct collapse and primordial BHs as potential progenitors. 

Aside from heavy seeding, a lone black hole can originate as a runaway from a triple black hole interaction in a galaxy nucleus \citemet{Saslaw1974,Volonteri2003}. The lightest black hole is ejected at about 1000~km~s$^{-1}$, and so at redshift 7 may travel up to $20$~arcsec in 100 Myr. We consider this scenario very unlikely as it compounds 2 unlikely events (being ejected and then coinciding with a cluster caustic). Searching the neighborhood of QSO1 for a suitable galaxy is complicated by the strong lensing, however, Ref. \citem{Furtak2023} do not report any nearby objects associated with QSO1. Either way, a quantitative examination into the origins of this object is beyond the scope of this paper.

\subsection*{Constraints on the gas mass}

As discussed above,
 contributions to the dynamical mass from ionized ISM gas, on scales of $\sim 100$~pc, are very small, and anyhow are taken into account. Contribution, on these scales from neutral gas, would make the upper limit on the stellar mass (based on dynamical modeling) even lower.

In the nuclear region, the amount of ionized gas is even  smaller. Indeed, as the emissivity scales quadratically with the density and the broad lines are coming from the BLR, which has a density higher than $10^9~cm^{-3}$, the mass of nuclear ionized gas inferred from the luminosity of the broad Balmer lines is only of order of a few solar masses  (see e.g. Ref.\citemet{Maiolino2024}), as also inferred from Equation \ref{eq:mion}.

The putative envelope of dense ionized gas that, according to the electron scattering scenarios, should produce the exponential line wings must also have a negligible mass. Indeed, such gas must recombine and its signature should be associated with one of the Balmer emission line components and, if the densities are as high as expected by those scenarios, the masses must be similarly low. Ref\citem{Rusakov2025} estimate an upper limit on the mass of the putative ionized envelope of $<10^5~M_\odot$ (for systems that are much more  luminous than QSO1), i.e. negligible relative to the mass estimated by us dynamically.

The envelope of neutral gas responsible for the putative Balmer scattering\citem{Naidu2025}\citemet{Chang2025} must also be negligible. This can be estimated by scaling from the case of the Rosetta Stone\citem{Juodzbalis2024b}, in which the column and density of neutral gas was estimated in detail, thanks to the presence of multiple transitions in absorption. In this case we obtain a mass of $\sim 6\times 10^4~M_\odot$, again negligible. 
In the case of the local LRD analogue
the ``Lord of LRDs''\citemet{Ji2025LRDlocal} or the
``Egg''\citemet{Lin2025LRDlocal}, which has a luminosity similar to QSO1, an upper limit on the mass of the neutral envelope of $7\times 10^5~M_\odot$ has been estimated - again negligible when compared to our dynamical mass estimates.

\clearpage

\section*{Data availability}
The data used in this study were obtained as part of JWST program ID 5015, and are
available from the \href{https://mast.stsci.edu/portal/Mashup/Clients/Mast/Portal.html}{Mikulski Archive for Space Telescopes} at the Space Telescope Science Institute, which is operated by the Association of Universities for Research in Astronomy, Inc., under NASA contract NAS 5-03127 for JWST.
The reduced data will be made available after review.




\bibliographymet{AGN}

\section*{Acknowledgements}

We are grateful to Priyamvada Natarajan for useful comments.
IJ acknowledges support by the Huo Family Foundation through a P.C. Ho PhD Studentship. 
This work is based on observations made with the National Aeronautics and Space Administration (NASA)/European Space Agency (ESA)/Canadian Space Agency (CSA) JWST. The data were obtained from the Mikulski Archive for Space Telescopes at the STScI, which is operated by the Association of Universities for Research in Astronomy, Inc., under NASA contract NAS 5-03127 for JWST. These observations are associated with programme PID 5015. RM, FD, JS, IJ, GJ acknowledge support from the Science and Technology Facilities Council (STFC), by the European Research Council (ERC) through Advanced Grant 695671 ``QUENCH'', by the UK Research and Innovation (UKRI) Frontier Research grant RISEandFALL. RM also acknowledges support from a Royal Society Research Professorship grant. SA acknowledges grant PID2021-127718NB-I00 funded by the Spanish Ministry of Science and Innovation/State Agency of Research (MICIN/AEI/ 10.13039/501100011033).
H\"U acknowledges funding by the European Union (ERC APEX, 101164796). Views and opinions expressed are however those of the authors only and do not necessarily reflect those of the European Union or the European Research Council Executive Agency. Neither the European Union nor the granting authority can be held responsible for them. AJB acknowledges funding from the "FirstGalaxies" Advanced Grant from the European Research Council (ERC) under the European Union’s Horizon 2020 research and innovation programme (Grant agreement No. 789056). SC acknowledges support by European Union’s HE ERC Starting Grant No. 101040227 - WINGS. SK has been supported by a Junior Research Fellowship from St Catharine's College, Cambridge and a Research Fellowship from the Royal Commission for the Exhibition of 1851. RS acknowledges support from the PRIN2022 MUR project 2022CB3PJ3 - First Light And Galaxy aSsembly (FLAGS) funded by the European Union - Next Generation EU and from the EU-Recovery Fund PNRR - National Centre for HPC, Big Data and Quantum Computing. VB, BL and SZ acknowledge the Texas Advanced Computing Center (TACC) for providing HPC resources under allocation AST23026. DS acknowledges support from the STFC, grant code ST/W000997/1. AT acknowledges support from the PRIN MUR project 2022935STW, funded by European Union-Next Generation EU, and from the INAF Fundamental Research 2023 Mini-grant project 'Cosmic Archaeology with the first black hole seeds'. GC acknowledges support from the INAF GO grant 2024 “A JWST/MIRI MIRACLE: MidIR Activity of Circumnuclear Line Emission”. KI acknowledges support from the National Natural Sci- ence Foundation of China (12233001), the National Key R\&D Program of China (2022YFF0503401), and the China Manned Space Program (CMS-CSST- 2025-A09). PGP-G acknowledges support from grant PID2022-139567NB-I00 funded by Spanish Ministerio de Ciencia e Innovaci\'on MCIN/AEI/10.13039/501100011033, FEDER {\it Una manera de hacer Europa}. MP acknowledges support through the grants PID2021-127718NB-I00, PID2024-159902NA-I00, and RYC2023-044853-I, funded by the Spain Ministry of Science and Innovation/State Agency of Research MCIN/AEI/10.13039/501100011033 and El Fondo Social Europeo Plus FSE+. RV acknowledges support from PRIN MUR "2022935STW" funded by European Union-Next Generation EU, Missione 4 Componente 2 CUP C53D23000950006 and from the Bando Ricerca Fondamentale INAF 2023, Theory Grant "Theoretical models for Black Holes Archaeology. JW gratefully acknowledges support from the Cosmic Dawn Center through the DAWN Fellowship. The Cosmic Dawn Center (DAWN) is funded by the Danish National Research Foundation under grant No. 140.

\newpage

\section*{Author information}

\subsection*{Affiliations}
\noindent
\hypertarget{inst:Kavli}$^{1}$Kavli Institute for Cosmology, University of Cambridge, Madingley Road, Cambridge CB3 0HA, UK
\\
\hypertarget{inst:Cav}$^{2}$Cavendish Laboratory, University of Cambridge, 19 JJ Thomson Avenue, Cambridge CB3 0HE, UK
\\
\hypertarget{inst:UniFlorence}$^{3}$
Universit\`a di Firenze, Dipartimento di Fisica e Astronomia, via G. Sansone 1, 50019 Sesto Fiorentino, Florence, Italy 2
\\
\hypertarget{inst:Arcetri}$^{4}$INAF – Arcetri Astrophysical Observatory, Largo E. Fermi 5, I50125, Florence, Italy
\\
\hypertarget{inst:UCL}$^{5}$Department of Physics and Astronomy, University College London, Gower Street, London WC1E 6BT, UK
\\
\hypertarget{inst:MPE}$^{6}$Max-Planck-Institut für extraterrestrische Physik, Gießenbachstraße 1, 85748 Garching, Germany
\\
\hypertarget{inst:CAB}$^{7}$Centro de Astrobiolog\'ia (CAB), CSIC–INTA, Cra. de Ajalvir Km.~4, 28850- Torrej\'on de Ardoz, Madrid, Spain
\\
\hypertarget{inst:CfA}$^{8}$Center for Astrophysics $|$ Harvard \& Smithsonian, 60 Garden St., Cambridge MA 02138, USA
\\
\hypertarget{inst:Texas}$^{9}$Department of Astronomy, University of Texas at Austin, Austin, TX 78712, USA
\\
\hypertarget{inst:Oxford}$^{10}$Department of Physics, University of Oxford, Denys Wilkinson Building, Keble Road, Oxford OX1 3RH, UK
\\
\hypertarget{inst:SNS}$^{11}$Scuola Normale Superiore, Piazza dei Cavalieri 7, I-56126 Pisa, Italy
\\
\hypertarget{inst:IAP}$^{12}$Sorbonne Universit\'e, CNRS, UMR 7095, Institut d'Astrophysique de Paris, 98 bis bd Arago, 75014 Paris, France
\\
\hypertarget{inst:Kapteyn}$^{13}$Kapteyn Astronomical Institute, University of Groningen, PO Box 800, 9700 AV Groningen, The Netherlands
\\
\hypertarget{inst:Toronto}$^{14}$Canadian Institute for Theoretical Astrophysics, 60 St George St, University of Toronto, Toronto, ON M5S 3H8, Canada
\\
\hypertarget{inst:Steward}$^{15}$Steward Observatory, University of Arizona, 933 North Cherry Avenue, Tucson, AZ 85721, USA
\\
\hypertarget{inst:IoA}$^{16}$Institute of Astronomy, University of Cambridge, Madingley Road, Cambridge, CB3 0HA, UK
\\
\hypertarget{inst:KIAA}$^{17}$Kavli Institute for Astronomy and Astrophysics, Peking University, Beijing 100871, China
\\
\hypertarget{inst:Waseda}$^{18}$Waseda Research Institute for Science and Engineering, Faculty of Science and Engineering, Waseda University, 3-4-1, Okubo, Shinjuku, Tokyo 169-8555, Japan
\\
\hypertarget{inst:Herts}$^{19}$Centre for Astrophysics Research, Department of Physics, Astronomy and Mathematics, University of Hertfordshire, College Lane, Hatfield, AL10 9AB, UK
\\
\hypertarget{inst:Catharine}$^{20}$St Catharine’s College, University of Cambridge, Trumpington Street, Cambridge CB2 1RL, UK
\\
\hypertarget{inst:Marseille}$^{21}$Aix Marseille Universit\'e, CNRS, CNES, LAM (Laboratoire d’Astrophysique de Marseille), UMR 7326, F-13388 Marseille, France
\\
\hypertarget{inst:Heidelberg}$^{22}$
Universit\"{a}t Heidelberg, Zentrum f\"{u}r Astronomie, Institut f\"{u}r Theoretische Astrophysik, D-69120 Heidelberg, Germany
\\
\hypertarget{inst:SCruz}$^{23}$Department of Astronomy and Astrophysics, University of California, Santa Cruz, 1156 High Street, Santa Cruz, CA 96054, USA 
\\
\hypertarget{inst:Sapienza}$^{24}$Dipartimento di Fisica, ‘Sapienza’ Universit\`
a di Roma, Piazzale Aldo Moro 2, I-00185 Roma, Italy
\\
\hypertarget{inst:Como}$^{25}$
Como Lake Center for Astrophysics, DiSAT, Universit\`
a degli Studi dell’Insubria, via Valleggio 11, 22100, Como, Italy
\\
\hypertarget{inst:OAR}$^{26}$INAF/Osservatorio Astronomico di Roma, Via di Frascati 33, I-00040 Monte Porzio Catone, Italy
\\
\hypertarget{inst:DAWN}$^{27}$Cosmic Dawn Center (DAWN), Copenhagen, Denmark
\\
\hypertarget{inst:NBI}$^{28}$Niels Bohr Institute, University of Copenhagen, Jagtvej 128, DK-2200, Copenhagen, Denmark
\\
\hypertarget{inst:Texas2}$^{29}$Department of Physics, University of Texas at Austin, Austin, TX 78712, USA
\\
\hypertarget{inst:Weinberg}$^{30}$Weinberg Institute for Theoretical Physics, Texas Center for Cosmology and Astroparticle Physics,
University of Texas at Austin, Austin, TX 78712, USA

\subsection*{Author contributions}
IJ led the data analysis and writing of the paper with CM performing independent verification of the results. Key contributions to the text were provided by FDE, RM, AM, HU, JS and XJ. Observations were prepared and data reduced by FDE, HU and MP. The remaining authors contributed to the interpretation of the results and provided comments on the manuscript.

\subsection*{Correspondence}

Correspondence should be addressed to I. Juodžbalis.

\section*{Ethics declarations}

\subsection*{Competing interests}

The authors declare no competing interests.




\bibliographysup{AGN}

\end{document}